# Minimizing Movement: Fixed-Parameter Tractability*

Erik D. Demaine[†‡]  MohammadTaghi Hajiaghayi[§‡¶]  Dániel Marx[‖**]

**Abstract**

We study an extensive class of movement minimization problems which arise from many practical scenarios but so far have little theoretical study. In general, these problems involve planning the coordinated motion of a collection of agents (representing robots, people, map labels, network messages, etc.) to achieve a global property in the network while minimizing the maximum or average movement (expended energy). The only previous theoretical results about this class of problems are about approximation, and mainly negative: many movement problems of interest have polynomial inapproximability. Given that the number of mobile agents is typically much smaller than the complexity of the environment, we turn to fixed-parameter tractability. We characterize the boundary between tractable and intractable movement problems in a very general setup: it turns out the complexity of the problem fundamentally depends on the treewidth of the minimal configurations. Thus the complexity of a particular problem can be determined by answering a purely combinatorial question. Using our general tools, we determine the complexity of several concrete problems and fortunately show that many movement problems of interest can be solved efficiently.

## 1 Introduction

In many applications, we have a relatively small number of mobile agents (e.g., a team of autonomous robots or people) moving cooperatively in a vast terrain or complex building to achieve some task. The number of cooperative agents is often small because of their expense: only small groups of people (e.g., emergency response or special police units) can effectively cooperate, and autonomous mobile robots are currently quite expensive (in contrast to, e.g., immobile sensors). Nonetheless, an accurate model of the immense/intricate environment they traverse, and their ability to communicate or otherwise interact (say, by limited-range wireless radios or walkie-talkies), is complicated and results in a large problem input. Thus, to compute the most energy-efficient motion in such a scenario, we allow the running time to be relatively large (exponential) in the number of agents, but it must be small (polynomial or even linear) in the complexity of the environment. This setup motivates the study of *fixed-parameter tractability* (FPT) [13, 16, 27, 21] for minimizing movement, with running time $f(k) \cdot n^{O(1)}$ for some function $f$, parameterized by the number $k$ of mobile agents.

A movement minimization problem is defined by a class of target configurations that we wish the mobile agents to form and a movement objective function. For example, we may wish to move the agents to

1. form a connected communication network (given a model of connectivity);

---

*A preliminary version of this paper appeared in *Proceedings of the 17th Annual European Symposium on Algorithms*, 2009.

[†]MIT Computer Science and Artificial Intelligence Laboratory, 32 Vassar St., Cambridge, MA 02139, USA, edemaine@mit.edu

[‡]Research supported in part by NSF grant CCF-1161626 and DARPA/AFOSR grant FA9550-12-1-0423.

[§]A. V. Williams Building, University of Maryland, College Park, MD 20742, USA, hajiagha@cs.umd.edu; and AT&T Labs — Research.

[¶]Research supported in part by NSF CAREER award 1053605, NSF grant CCF-1161626, ONR YIP award N000141110662, DARPA/AFOSR grant FA9550-12-1-0423, and a University of Maryland Research and Scholarship Award (RASA).

[‖]Computer and Automation Research Institute, Hungarian Academy of Sciences (MTA SZTAKI), Budapest, Hungary, dmarx@cs.bme.hu

[**]Research supported by the European Research Council (ERC) grant "PARAMTIGHT: Parameterized complexity and the search for tight complexity results," reference 280152.

2. form a fault-tolerant (say, $k$-connected) communication network;

3. disperse throughout the environment (forming an independent set in a graph representing proximity, which also has applications to map labeling [12, 22, 32, 23, 11]).

4. collect into a small number of collocated groups (e.g., to form teams or arrange for a small number of deliveries);

5. form a perfect matching of communication pairs (e.g., to exchange information in each step of a network multicast);

6. arrange into a desired topological formation such as a grid (a common goal in providing reliable communication infrastructure);

7. service a collection of clients (e.g., sensors, who may themselves be mobile);

8. separate "main" agents (say, representing population) from "obnoxious" agents (say, representing power plants); or

9. augment an existing immobile network to achieve a desired property such as connectivity (viewing the agents as starting at infinity, and thus minimizing the number of moved/used resources as in [4, 6, 7]).

This list is just a partial collection of interesting agent formations; there are many other desiderata of practical interest, including combinations of different constraints. See Section 3.1 for more formal examples of problems and how our theory applies to them.

In the general formulation of the movement problem, we are given an arbitrary metric defining feasible motion, a graph defining "connectivity" (possibly according to the infinite Euclidean plane), and a desired property of the connectivity among the agents defined by a class $\mathcal{G}$ of graphs. We view the agents as "pebbles" located at vertices of the connectivity graph (and we use the two terms interchangeably). Our goal is to move the agents so that they induce a subgraph of the connectivity graph that possesses the desired property, that is, belongs to the class $\mathcal{G}$. There are three natural measures of agent motion that we might want to minimize: the total amount of motion, the maximum motion of any agent, and the number of moved agents. To obtain further generality and to model a wider range of problems, we augment this model with additional features: the agents have types, desired solutions can require certain types of agents, multiple agents can be located at the same vertex, and the cost of the movement can be different (even nonmetric) for the different agents.

To what level of generality can we solve these movement problems? Several versions have been studied from an approximation algorithms perspective [11, 18], in addition to various specific problems considered less formally in practical scenarios [4, 6, 7, 20, 26, 29, 31, 12, 22, 32, 23]. Unfortunately, most forms of the movement problem are NP-complete, and furthermore are often hard to approximate even within polynomial factors [11]. Nonetheless, the problems are of significant practical interest, and the motion must be kept small in order to minimize energy consumption. Fortunately, as motivated above, the number of mobile agents is often small. Thus we have a natural context for considering fixed-parameter algorithms, i.e., algorithms with running time $f(k) \cdot n^{O(1)}$, where parameter $k$ is the number of mobile agents.

In this paper, we develop general efficient fixed-parameter algorithms for a broad family of movement problems. Furthermore, we show our results are tight by characterizing, in a very general setting, the line between fixed-parameter tractability and intractability. It turns out that the notion of treewidth plays an important role in defining this boundary line. Specifically we show that, for problems closed under edge addition (i.e., adding an edge to the connectivity graph cannot destroy a solution), the complexity of the problem depends solely on whether the edge-deletion minimal graphs of the property have bounded treewidth. If they all have bounded treewidth, we show how to solve a very general formulation of the problem with an efficient fixed-parameter algorithm. If they have unbounded treewidth, we show that even very simple questions are W[1]-hard, meaning there is no efficient fixed-parameter algorithm under the standard parameterized complexity assumption FPT $\neq$ W[1]. (This assumption is the parameterized analog of P $\neq$ NP: it is stronger than P $\neq$ NP, but weaker than the Exponential Time Hypothesis.)

Our framework for movement problems is very general, and sometimes this full generality is unnecessary. Thus, we begin in Section 2 with a simplified version of our framework, and describe several of its applications to specific movement problems in Section 2.1. Then, Section 3 presents the general version of our framework, which allows multiple types of overlapping agents, and Section 3.1 presents many further applications of this framework to specific movement problems. Finally, Section 4 presents further improvements for specific problems and for specific graph classes such as planar graphs. The formal definition of all the concepts appear in Section 5. The results are proved in Sections 6–8.

## 2 Simplified Results

We start by presenting simplified versions of our main results, which handle only a simpler formulation of the movement problem, but are already capable of determining the complexity of several natural problems. The full model is presented in Sections 3– 4 and the formal definitions can be found in Section 5.

A (simplified) movement problem is specified by a *graph property*: an (infinite) set $\mathcal{G}$ of desired configurations. Given a graph $G$ with $k$ agents on the vertices, the task in the *movement problem* is to move the agents to $k$ distinct vertices such that the graph induced by the $k$ agents is in $\mathcal{G}$. The goal is to minimize the "cost" of the movements, such as the total number of steps the agents move, the maximum movement of an agent, or the number of agents that move at all.

In fact, we define the *cost* of a solution to be the sum of costs of each agent's movement, where we are given a (polynomially computable) *movement cost function* for each agent specifying a nonnegative integer cost of moving that agent to each vertex in the graph. This definition obviously includes counting the total number of steps agents move as a special case, as well as modeling nonmetric terrains, agents of different speeds, immobile agents, regions impassable by certain agents, etc. This definition of movement cost also includes the other objectives mentioned above as special cases. To minimize the number of moved agents, we can specify a movement cost function for each agent of 0 to remain stationary and 1 to make any move. To minimize the maximum motion of an agent, we can binary search on the maximum movement cost $\tau$, and modify the movement cost function to jump to $\infty$ whenever exceeding this threshold $\tau$.

Our algorithmic result for these (simplified) movement problems considers graph properties that are closed under edge addition (which holds in particular for properties that model some notion of connectivity):

**Theorem 2.1.** *If $\mathcal{G}$ is a decidable graph property that is closed under edge addition, and the edge-deletion minimal graphs in $\mathcal{G}$ have bounded treewidth, then the movement problem can be solved in time $f(k) \cdot n^{O(1)}$.*

We prove a matching hardness result for Theorem 2.1: if the edge-deletion minimal graphs in $\mathcal{G}$ have unbounded treewidth, then it is hard to answer even some very simple questions.

**Theorem 2.2.** *If $\mathcal{G}$ is any graph property that is closed under edge addition and has unbounded treewidth, then the movement problem is W[1]-hard parameterized by $k$, already in the special case where each agent is allowed to move at most one step in the graph.*

Theorems 2.1 and 2.2 show that the algorithmic/complexity question of whether a given movement problem is FPT can be reduced to the purely combinatorial question of whether a certain set of graphs has bounded treewidth. Thus treewidth plays an essential role in the complexity of the problem, which is not apparent at first sight. As we shall see in the examples below, this connection with treewidth allows us to understand how subtle differences in the definition of the problem (e.g., connectivity vs. 2-connectivity or edge-disjoint paths vs. vertex-disjoint paths) change the complexity of the problem.

Theorems 2.1 and 2.2 considered properties closed under edge addition. We prove another general result, which considers *hereditary* properties, i.e., properties closed under taking induced subgraphs:

**Theorem 2.3.** *Let $\mathcal{G}$ be a decidable hereditary property. If $\mathcal{G}$ does not contain all cliques or does not contain all independent sets, then the movement problem is W[1]-hard parameterized by $k$, already in the special case where each agent is allowed to move at most one step in the graph.*

### 2.1 Applications of Simplified Results

Theorems 2.1–2.3 immediately characterize the complexity of several natural problems:

**Example: CONNECTIVITY** Move the pebbles (agents) so that they are connected and on distinct vertices. The parameter is the number $k$ of pebbles. Now $\mathcal{G}$ contains all connected graphs. Clearly, $\mathcal{G}$ is closed under edge addition and the edge-deletion minimal graphs are trees. Trees have treewidth 1, hence by Theorem 2.1, this movement problem is fixed-parameter tractable for any movement cost function. □

**Example: 2-CONNECTIVITY** Move the pebbles so that they induce a 2-connected graph and the pebbles are on distinct vertices. The parameter is the number $k$ of pebbles. Now $\mathcal{G}$ contains all 2-connected graphs and clearly $\mathcal{G}$ is closed under edge addition. The edge-deletion minimal graphs have unbounded treewidth: subdividing every edge of a clique gives an edge-deletion-minimal 2-connected graph. Thus by Theorem 2.2, it is W[1]-hard to decide whether there is a solution where each pebble moves at most one step. □

**Example: GRID** Move the $k$ pebbles so that they are on distinct vertices and they form a $\lfloor\sqrt{k}\rfloor \times \lfloor\sqrt{k}\rfloor$ square grid. The parameter is the number $k$ of pebbles. Let $\mathcal{G}$ contain all graphs containing a spanning square grid subgraph. Clearly, $\mathcal{G}$ is closed under edge addition and the edge-deletion minimal graphs are grids, which have arbitrarily large treewidth. Thus Theorem 2.2 implies that it is W[1]-hard, to decide whether there is a solution where each pebble moves at most one step. □

**Example: MATCHING** Move the pebbles so that the pebbles are on distinct vertices and there is a perfect matching in the graph induced by the pebbles. The parameter is the number of pebbles. Let $\mathcal{G}$ contain all graphs that have a perfect matching. The edge-deletion minimal graphs are perfect matchings, (i.e., $k/2$ independent edges on $k$ vertices), so they have treewidth 1. By Theorem 2.1, the movement problem is FPT. □

**Example: DISPERSION** Move the pebbles to distinct vertices and such that no two pebbles are adjacent. The parameter is the number $k$ of pebbles. Here $\mathcal{G}$ contains all independent sets. Because $\mathcal{G}$ is hereditary and the maximum clique size is 1, Theorem 2.3 implies that the movement problem is W[1]-hard, even in the case when each pebble is allowed to move at most one step. □

## 3 Main Results

In this section, we present the full generality of the problem we consider and results we obtain (note that the formal definitions are collected in Section 5). In particular, this generalization removes several limitations of the simplified version presented above, informally summarized as follows:

1. In many cases agents have different types (e.g., some of the agents are servers, some are clients, etc.). and the solution should take these types into account.

2. If, for example, the task is to provide connectivity between two specific vertices $s$ and $t$, then the model should be capable of specifying these two distinguished vertices in the input.

3. Agents should be able to share vertices, i.e., we should not require that the agents move to distinct vertices. In fact, we may require that more than one agent is moved to a single vertex, e.g., if the task is to move a server to each client.

4. The graph induced by the agents might not suffice to certify that the solution is correct (e.g., if the requirement is that the agents are at distance-2 from each other). We might want to include (a bounded number of) unoccupied vertices into the solution in order to produce a witness showing that the agents have the correct configuration.

5. In some scenarios, agents are divided into "clients" that need to be satisfied somehow and "facilities" that are helpful for satisfying the clients but otherwise do not introduce any additional constraints to the problem. In many cases, we are able to extend the fixed-parameter tractability results such that the parameter is the number of client agents only, and thus the number of facility agents can be arbitrarily large. We introduce a similar generalization with an unbounded number of "obnoxious" agents which can interfere with clients, but otherwise do not introduce any requirements on their own.

Formally, the general model we consider divides the agents into three types—client, facility, and obnoxious agents—and the parameter is just the number of clients, which can be much smaller than the total number of agents. The clients can require collocated or nearby *facility agents*, among a potentially large set of facility agents, which themselves are mobile. Intuitively, facilities provide some service needed by clients. Clients can also require at most a certain number (e.g., zero) of collocated *obnoxious agents* (again among a potentially large, mobile set), which can represent dangerous or undesirable resources. In other words, adding facility agents or removing obnoxious agents does not make a correct solution invalid. More generally, there can be many different subtypes of client, facility, and obnoxious agents, and we may require a particular pattern of these types.

A (general) movement problem specifies a *multicolored graph property*: an (infinite) set $\mathcal{G}$ of desired configurations, each specifying a desired subgraph and how that subgraph should be populated by different types of agents (a *multicolored graph*). Each agent type (color) is specified as client, facility, or obnoxious, but there can be more than three types; in this way, we can specify different types of client agents that need to interact in a particular way, or need particular types of nearby facility agents. The goal of the *movement problem* is to move the agents into a configuration containing at most $\ell$ vertices that contain all $k$ client agents and induce a "good" target pattern. A good target pattern is a multicolored graph that is either in the set $\mathcal{G}$ or it "dominates" some multicolored graph $G \in \mathcal{G}$ in the sense that it contains more facility agents and fewer obnoxious agents of each color at each vertex. To emphasize that the goal is to create a pattern containing all the client agents (which may contain only a subset of facility and obnoxious agents), we will sometimes call the client agents "main agents" and use the two terms interchangeably.

A mild technical condition that we require is that the multicolored graph property $\mathcal{G}$ is *regular*: for every fixed numbers $k$ and $\ell$, there are only finitely many graphs in $\mathcal{G}$ with at most $\ell$ vertices and at most $k$ client agents (as we do not bound the number of obnoxious and facility agents here, this is a nontrivial restriction). In other words, there should be only finitely many minimal ways to satisfy a bounded number of clients in a bounded subgraph. For example, the property requiring that the number of facility agents at each vertex is not less than the number of obnoxious agents at that vertex is *not* a regular property. Note that this restriction does not say that there is only a finite number of good configurations, it only says that there is a finite number of *minimal* good configurations: as mentioned in the previous paragraph, we allow configurations having any number of extra facility agents.

For a regular multicolored graph property, the corresponding movement problem is as follows: given an initial configuration (a multicolored graph), to minimize the total cost of all movement subject to reaching one of the desired target configurations in $\mathcal{G}$ with at most $\ell$ vertices, where both $\ell$ and the number $k$ of client agents are parameters. As before, we are given a movement cost function for each agent, an arbitrary (polynomially computable) function specifying the nonnegative integer cost of moving that agent to each vertex in the graph.

Our main algorithmic result considers properties that are closed under edge addition (for example, properties that model some notion of connectivity). Besides requiring that the graph property is regular, another mild technical assumption is that two obnoxious agents of the same type behave similarly, i.e., the cost of moving them from vertex $v_1$ to $v_2$ has the same cost.

**Theorem 3.1.** *If $\mathcal{G}$ is a regular multicolored graph property that is closed under edge addition, and if the edge-deletion minimal graphs in $\mathcal{G}$ have bounded treewidth, then the movement problem can be solved in $f(k, \ell) \cdot n^{O(1)}$ time, assuming that the movement cost function is the same on any two agents of the same obnoxious type that are initially located on the same vertex.*

Our main algorithm (Section 6) uses several tools from fixed-parameter tractability, color coding, and graph structure theory, in particular treewidth. This combination of techniques seems interesting in its own right.

We prove in Section 7 a matching hardness result for Theorem 3.1: if the edge-deletion minimal graphs in $\mathcal{G}$ have unbounded treewidth, then it is hard to answer even some very simple questions. Thus treewidth plays an essential role in the complexity of the problem, which is not apparent at first sight.

**Theorem 3.2.** *If $\mathcal{G}$ is any (possibly regular) multicolored graph property that is closed under edge addition, and for every $w \geq 1$, there is an edge-deletion minimal graph $G_w \in \mathcal{G}$ with treewidth at least $w$ and at least one client agent on each vertex (but no other type of agent), then the movement problem is W[1]-hard with the combined parameter $(k, \ell)$, already in the special case where each agent is allowed to move at most one step.*

If a movement problem can be modeled with colored pebbles and the target patterns are closed under edge addition, then the complexity of the problem can be determined by solving the (sometimes nontrivial) combinatorial question of whether the minimal configurations have bounded treewidth. The minimal configurations are those pebbled graphs that are acceptable solutions, but removing any edge makes them unacceptable.

As before, we also obtain a general hardness result for multicolored graph properties that are not closed under edge addition, but rather are hereditary, i.e., closed under taking induced subgraphs:

**Theorem 3.3.** *Let $\mathcal{G}$ be a hereditary property where each vertex has exactly one client agent and there are no other type of pebbles. If $\mathcal{G}$ does not contain all cliques or does not contain all independent sets, then the movement problem is W[1]-hard with the combined parameter $(k, \ell)$, already in the special case where each agent is allowed to move at most one step in the graph.*

The proof of Theorem 3.3 (in Section 8.6) uses a hardness result by Khot and Raman [24] on the parameterized complexity of finding induced subgraphs with hereditary properties.

## 3.1 Applications of Main Results

Theorems 3.1 and 3.2 characterize the complexity of several additional natural problems beyond Section 2.1:

**Example: CONNECTIVITY (collocation allowed)** The connectivity problem discussed in Section 2.1 required that all the pebbles are moved to distinct vertices. For example, moving all the pebbles to the same vertex is not a correct solution. It could be however that some applications are more faithfully expressed if we allow pebbles to share vertices. It is easy to express this variant using Theorem 3.1. Let $\mathcal{G}$ contain all connected graphs with *at least* one pebble on each vertex. Setting $\ell = k$, it follows from Theorem 3.1 that this variant of the problem is FPT parameterized by $k$. □

**Example: $s$-$t$ CONNECTIVITY (few pebbles)** Move the pebbles to form a path of pebbled vertices between fixed vertices $s$ and $t$. The parameter is the number $k$ of pebbles. Now there are two main colors of pebbles, call them red and blue, and $\mathcal{G}$ consists of all graphs containing exactly two red pebbles and a path between them using only vertices with blue pebbles. We reduce $s$-$t$ CONNECTIVITY to this movement problem by putting red pebbles at $s$ and $t$, and giving them an infinite movement cost to any other vertices. Clearly, $\mathcal{G}$ is closed under edge addition and the edge-deletion minimal graphs are paths. Paths have treewidth 1, so by Theorem 3.1, this problem is fixed-parameter tractable. □

In the next example, we show that a much more general version of $s$-$t$ CONNECTIVITY is FPT: instead of parameterizing by the number $k$ of pebbles, we can parameterize by the maximum length $L$ of the path. Thus we can have arbitrarily many pebbles that might form the path, and allow the runtime to be exponential in the length of the path.

**Example: $s$-$t$ CONNECTIVITY (bounded length)** Move the pebbles to form a path of pebbled vertices of length at most $L$ between fixed vertices $s$ and $t$. The parameter is the length $L$. Now we define one main color of pebbles, red, and one facility color of pebbles, blue, and we define $\mathcal{G}$ as in the previous example. Again by Theorem 3.1, this problem is fixed-parameter tractable in the combined parameter $(k, \ell)$; in the example, we have $k = 2$ and $\ell = L + 1$. □

**Example: STEINER CONNECTIVITY** Connect the red pebbles (representing terminals) by moving the blue pebbles to form a Steiner tree. The parameter is the number of red pebbles plus the number of blue pebbles in the *solution* Steiner tree. This is simply a generalization of *s-t* CONNECTIVITY to more than two red pebbles. Again by Theorem 3.1 the problem is fixed-parameter tractable with this parameterization (the edge-deletion minimal graphs are trees), even when the number of blue pebbles is very large in the input. □

**Example: *s-t* *d*-CONNECTIVITY (fixed *d*)** Move the pebbles so that there are $d$ vertex-disjoint paths using pebbled vertices between two fixed vertices $s$ and $t$. The parameter is the total length $L$ of the $d$ paths in the solution. Now we use one main color, red, and one facility color, blue, and $\mathcal{G}_d$ consists of all graphs containing two vertices with a red pebble on each, and having $d$ internally vertex-disjoint paths between these two vertices, with blue pebbles on each internal vertex. In the input instance, there are red pebbles on $s$ and $t$, and the cost of moving them is infinite. Clearly, $\mathcal{G}_d$ is closed under edge addition and the edge-deletion minimal graphs are series-parallel (as they consist of $d$ internally vertex disjoint paths connecting two vertices), which have treewidth 2. Hence, by Theorem 3.1, this movement problem is fixed-parameter tractable with respect to $L$, for every fixed $d$. Again the number of blue pebbles can be arbitrarily large. □

The previous example shows that *s-t d*-CONNECTIVITY is FPT for every fixed value of $d$, i.e., for every fixed $d$, there is an $f(L) \cdot n^{O(1)}$ time algorithm. However, this statement does not make it clear if the degree of $n$ depends on $d$ or not. To show that the degree of $n$ is independent of $d$ and the problem can be solved in time $f(L,d) \cdot n^{O(1)}$, we need to encode the number $d$ in the input of the movement problem. We use dummy green pebbles for this purpose.

**Example: *s-t* *d*-CONNECTIVITY (unbounded version)** Move the pebbles so that there are $d$ vertex-disjoint paths using pebbled vertices between two fixed vertices $s$ and $t$, where $d$ is a number given in the input. The parameter is the total length $L$ of the solution paths. First, if $d$ is larger than the bound on the total length of the paths, then there is no solution. Otherwise, we can assume $d$ is a fixed parameter. Now we use two main colors, red and green, and one facility color, blue. A graph $G$ is in $\mathcal{G}$ if the blue pebbles form $d$ internally vertex-disjoint paths between two vertices containing red pebbles, where $d$ is the number of green pebbles in $G$. Thus we use green pebbles to "label" a graph $G$ in $\mathcal{G}$ according to what level of connectivity it attains. Again $\mathcal{G}$ is closed under edge addition and the edge-deletion minimal graphs are series-parallel, which have treewidth 2, so by Theorem 3.1, the movement problem is fixed-parameter tractable with respect to $k := 2$ and $\ell := L$. In the initial configuration, we put red pebbles on $s$ and $t$ with infinite movement cost, and we place $d$ green pebbles arbitrarily in the graph. The target configuration we obtain will have exactly $d$ green pebbles, and thus $d$ vertex-disjoint paths, because these are main pebbles. □

We can also consider the edge-disjoint version of *s-t* connectivity. We need the following combinatorial lemma to characterize the minimal graphs:

**Lemma 3.4.** *Let $G$ be a connected graph and assume that there are $d$ edge-disjoint paths between vertices $s$ and $t$ in $G$, but for any edge $e \in E(G)$, there are at most $d - 1$ edge-disjoint paths between $s$ and $t$ in $G \setminus e$. Then the treewidth of $G$ is at most $2d + 1$.*

*Proof.* Note that the size of the minimum $s - t$ cut is exactly $d$. We use the folklore observation that there is a noncrossing family of minimum $s - t$ cuts covering every minimum $s - t$ cut. (We say that two $s - t$ cuts $C_1, C_2 \subseteq E(G)$ *cross* if there is a vertex $v_1$ reachable from $s$ in $G \setminus C_1$ but not in $G \setminus C_2$, and there is a vertex $v_2$ reachable from $s$ in $G \setminus C_2$ but not in $G \setminus C_1$.) The formal statement that we use is that there is a sequence $\{s\} \subseteq X_1 \subseteq X_2 \subseteq \cdots \subseteq X_r \subseteq V(G) \setminus \{t\}$ such that (1) for every $1 \leq i \leq t$, exactly $d$ edges go between $X_i$ and $V(G) \setminus X_i$, and (2) if edge $e$ appears in a minimum $s - t$ cut, then there is an $1 \leq i \leq r$ such that $e$ connects $X_i$ and $V(G) \setminus X_i$. In our case, every edge appears in a minimum $s - t$ cut, thus the edges leaving the $X_i$'s cover every edge.

Let $Y_i$ be the endpoints of the edges connecting $X_i$ and $V(G) \setminus X_i$. Let $T$ be a tree that is a path with nodes $v_1, \ldots, v_r$ and let us define $B_i := Y_i \cup \{s, t\}$; clearly $|B_i| \leq 2d + 2$. We claim that $(T, B_i)$ is a tree decomposition of width $2d + 1$. From our discussion above, it is clear that every edge appears in one of the bags. To see the connectedness property, suppose that $v \in B_i$ and $v \notin B_j$ for some $j > i$. We need to show that $v \notin B_{j'}$ for any $j' > j$. As $v \in B_j$, vertex $v$ is the endpoint of an edge leaving $X_i$. Thus either $v \in X_i$ or $v$ is adjacent to a vertex of $X_i$. In both cases, we have $v \in X_j$: in the first case, this follows from $X_i \subseteq X_j$; in the second case, $v \notin X_j$ would mean that the edge connecting $v$ with $X_i$ does not leave $X_j$, which is only possible if this edge is contained in $X_j$. Thus $v \in X_j$ and $v$ has no neighbor outside $X_{j'}$, hence $X_j \subseteq X_{j'}$ implies that $v$ is in $X_{j'}$ as well and has no neighbor outside $X_{j'}$, that is, $v \notin X_{j'}$. □

**Example:** *s*-*t* *d*-**EDGE-CONNECTIVITY (fixed *d*)** Move the pebbles so that there are $d$ edge-disjoint paths of pebbled vertices between $s$ and $t$. The parameter is the total length $L$ of the paths. Now we use one main color, red, and one facility color, blue, and $\mathcal{G}_d$ contains all graphs containing two vertices with a red pebble on each and having $d$ edge-disjoint paths between these two vertices, with blue pebbles on each path vertex. By Lemma 3.4, the edge-deletion minimal graphs have treewidth $O(d)$. Hence, by Theorem 3.1, the movement problem is fixed-parameter tractable with respect to $L$. □

The previous example shows that *s*-*t* *d*-EDGE-CONNECTIVITY is FPT parameterized by $L$ for every fixed value of $d$. As in the vertex-disjoint, we can ask if the problem is FPT if $d$ is part of the input and the parameters are $L$ and $d$. Somewhat surprisingly, unlike the vertex-disjoint case, the problem becomes hard:

**Example:** *s*-*t* *d*-**EDGE-CONNECTIVITY (unbounded version)** Move the pebbles so that there are $d$ edge-disjoint paths of pebbled vertices between $s$ and $t$, where $d$ is a number given in the input. We use three main colors: red, green, and blue. A graph $G$ is in $\mathcal{G}$ if the blue pebbles form $d$ edge-disjoint paths between two vertices containing red pebbles, where $d$ is the number of green pebbles in $G$. We show that $\mathcal{G}$ contains edge-deletion minimal graphs of arbitrary large treewidth, so by Theorem 3.2, it is W[1]-hard to decide whether there is a solution where each of the $k$ pebbles move at most one step each. Assume $d$ is even and let $G$ be a graph consisting of vertices $s$, $t$, and $d$ vertex-disjoint $s-t$ paths of length $d + 2$ such that vertices $v_{i,0}, \ldots, v_{i,d+1}$ are the internal vertices of the $i$th path. Now for every odd $i$ and odd $1 \leq j < d$, let us identify vertices $v_{i,j}$ and $v_{i+1,j}$, and for every even $i < d$ and even $1 < j \leq d$, let us identify $v_{i,j}$ and $v_{i+1,j}$ (see Figure 1). There are $d$ edge-disjoint $s$-$t$ paths in this graph, but there are at most $d - 1$ such paths after the deletion of every edge. (It is easy to see that every edge is in an $s$-$t$ cut of exactly $d$ edges.) Thus $G$ is an edge-deletion minimal member of $\mathcal{G}$. Furthermore, the graph contains a $d/2 \times d/2$ grid, so the treewidth is $\Omega(d)$. □

**Example: FACILITY LOCATION (collocation version)** Move client and facility agents so that each client agent is collocated with at least one facility agent and the client agents are at distinct locations. The parameter is the number of client agents. We use one main color, red, for the clients, and one facility color, blue, for the facilities, and $\mathcal{G}$ contains all graphs in which every vertex contains exactly one red and one blue pebble. The edge-deletion minimal graphs in $\mathcal{G}$ have no edges, so have treewidth 0. By Theorem 3.1, the movement problem is fixed-parameter tractable parameterized by the number of main pebbles, i.e., the number of client agents. The number of facilities can be unbounded, which is useful, e.g., to organize a small team within a large infrastructure of wired network hubs or mobile satellites. □

**Example: FACILITY LOCATION (distance-*d* version)** Move client and facility agents so that each client agent is within distance at most $d$ from at least one facility pebble and the client agents are at distinct locations. Now we use two main colors, red and green, and one facility color, blue. Let $\mathcal{G}$ contain all graphs that contain some number $d$ of green pebbles and each red pebble is at distance at most $d$ from some blue pebble. Given a graph with $k$ main (red) pebbles and some number of facility (blue) pebbles, we add $d$ dummy green pebbles and ask whether there is a solution on $\ell := k(d + 1) + d$ vertices. If we move the pebbles so that each red pebble is at distance $d$ from some blue pebble, then there are $k(d + 1) + d$ vertices that contain all $d$ of the green pebbles and induce a graph that belongs to $\mathcal{G}$ (such a set can be obtained by taking all the red and green pebbles and selecting, for each red pebble, a path of at most $d$ additional vertices that connect it to a blue pebble). We claim that the edge-deletion minimal graphs in $\mathcal{G}$ are forests, and hence have treewidth 1. Consider an edge-deletion minimal graph $G \in \mathcal{G}$, and for each vertex $v$ without a blue pebble, select an edge $uv$ that goes to a neighbor $u$ that is closer to some blue pebble than $v$. If an edge is not selected in this process, then it can be removed (it does not change the distance to the blue pebbles), so by the minimality of $G$, every edge is selected. Each connected component contains at least one blue pebble. This means that, in each connected component, the number of selected edges is strictly smaller than the number of vertices, i.e., each component is a tree. Thus, by Theorem 3.1, the movement problem is FPT. □

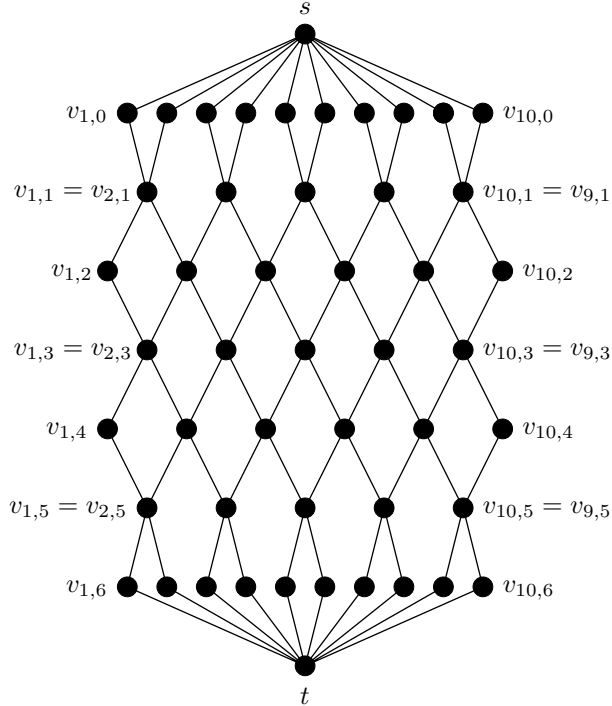

Figure 1: The graph $G$ for $d = 5$ in the discussion of *s-t* $d$-EDGE-CONNECTIVITY (unbounded version).

On the other hand, FACILITY LOCATION becomes W[2]-hard if the parameter is the number of facilities, while the number of clients can be unbounded (Theorem 3.5 below). This result cannot be obtained using the general result of Theorem 3.2 because in this statement the parameter is the number of facility pebbles. However, it is not difficult to give a problem-specific hardness proof for this variant.

**Theorem 3.5.** *For every fixed $d \geq 0$, FACILITY LOCATION (distance $d$ version) is W[2]-hard parameterized by the number of facilities, even if each pebble is allowed to move at most one step in the graph.*

*Proof.* To show that the problem is W[2]-hard, we show a reduction from MINIMUM DOMINATING SET (recall that a set $S \subseteq V(G)$ is a *dominating set* of $G$ if every vertex of $G$ is either in $S$ or adjacent to a vertex in $S$). Given a graph $G$ and an integer $k$, we construct an instance of FACILITY LOCATION with $k$ facility pebbles which can be solved by moving each pebble at most one step if and only if $G$ has a dominating set of size $k$. Let $v_1, \ldots, v_n$ be the vertices of $G$. We construct a graph $F$ as follows. We start with vertices $s$, $b_1, \ldots, b_n, c_1, \ldots, c_n$, where $s$ is connected to every $b_i$. If $v_i$ and $v_j$ are neighbors in $G$, then $b_i$ and $c_j$ are connected with a path having $d$ internal vertices. We place $k$ facility pebbles on $s$ and one main pebble on each $c_i$.

If $G$ has a dominating set $v_{i_1}, \ldots, v_{i_k}$, then we move the $k$ facility pebbles to $b_{i_1}, \ldots, b_{i_k}$, and if vertex $v_j$ of $G$ is dominated by its neighbor $v_{i_\ell}$, then we move the main pebble at $c_j$ one step closer to $b_{i_\ell}$. It is clear that each main pebble will be at distance exactly $d$ from some facility pebble. The other direction is also easy to see: if the facility pebbles move to vertices $b_{i_1}, \ldots, b_{i_k}$, then vertices $v_{i_1}, \ldots, v_{i_k}$ form a dominating set in $G$. □

We remark that the fixed-parameter tractability of facility location problems has been investigated in [15]. The model studied there is somewhat different from the one studied here, but [15, Theorem 6] gives a very similar simple reduction from (essentially) DOMINATING SET to facility location parameterized by the number of facilities.

**Example: SEPARATION**  Move client agents (say, representing population) and/or obnoxious agents (say, representing power plants) so that each client agent is collocated with at most $o$ obnoxious pebbles. The parameter is the number of client agents. Here $\mathcal{G}$ contains all graphs with the desired bounds, so the edge-deletion minimal graphs have no edges, which have treewidth 0. By Theorem 3.1, the movement problem is fixed-parameter tractable. As in previous examples, we can make $o$ an input to the problem. □

# 4 Further Results

In addition to our general classification and specific examples, we present many additional fixed-parameter results. These results capture situations where the general classification cannot be applied directly, or the general results apply but problem-specific approaches enable more efficient algorithms. Specifically, we consider situations where the graphs are more specific (e.g., almost planar), the property is not closed under edge addition, or the number of client agents is not bounded. Our aim is to demonstrate that there are many problem variants that can be explored and that there is a vast array of algorithmic techniques that become relevant when studying movement problems. In particular, results from algorithmic graph minor theory (Section 8.1), Courcelle's Theorem (Section 8.2), bidimensionality (Section 8.2), the fast set convolution algorithm of Björklund et al. (Section 8.3), and Canny's Roadmap Algorithm (Section 8.4) all find uses in this framework.

## 4.1 Planar Graphs and $H$-Minor-Free Graphs

Our general characterization makes no assumptions on the connectivity structure: it is an arbitrary graph. However, significantly stronger results can be achieved if we have some restriction on the connectivity graph. For example, many road networks, fiber networks, and building floorplans can be accurately represented by planar graphs. We show that, for planar graphs, the fixed-parameter algorithms of Theorem 3.1 work even if we remove the requirement that $\mathcal{G}$ is closed under edge addition. That is, we can express for example that the pebbles induce an independent set.

In many cases, approximation and fixed-parameter tractability results for planar graphs generalize to arbitrary surfaces, to graphs of bounded local treewidth, and to $H$-minor-free graph classes. These generalizations are made possible by the algorithmic consequences of the Graph Minor Theorem [10]; see Section 8.1. To obtain maximum generality, we state the result on planar graphs generalized to arbitrary $H$-minor-free classes:

**Theorem 4.1.** *If $\mathcal{G}$ is a regular multicolored graph property, then for every fixed graph $H$, the movement problem can be solved on $H$-minor-free graphs in $f(k,\ell) \cdot n^{O(1)}$ time, assuming that the movement cost function is the same on any two agents of the same obnoxious type that are initially located on the same vertex.*

We stress that in Theorem 4.1, unlike in Theorem 3.1, the property $\mathcal{G}$ is *not* required to be closed under edge addition. One possible application scenario where these generalizations of planar graphs play a role is the following. The terrain is a multi-level building, where the connectivity graph is planar on each level, and there are at most $d$ connections between two adjacent levels (for some fixed $d \geq 4$). It is easy to see that the graph is $K_{d+1}$-free: a $K_{d+1}$ minor would be contained on one level. Thus, for every fixed value of $d$, Theorem 4.1 applies for such connectivity graphs.

We also consider two specific problems in the context of planar graphs.

## 4.2 Bidimensionality

In Section 8.2, we show that bidimensionality theory can be exploited to obtain algorithms for movement problems on planar graphs. In particular, we show that the version of DISPERSION (see Section 2.1) where each pebble can move at most one step admits a subexponential parameterized algorithm. The proof uses a combination of bidimensionality theory, parameter-treewidth bounds, grid-minor theorems, Courcelle's Theorem, and monadic second-order logic.

### 4.3 Planar STEINER CONNECTIVITY

In the STEINER CONNECTIVITY problem (see Section 3.1), the goal is to connect one type of agents ("terminals") using another type of agents ("connectors"). Our general characterization shows that this problem is fixed-parameter tractable if the numbers of both types of agents are bounded, while it becomes W[1]-hard if only the number of connector agents is bounded and the number of terminal pebbles is unbounded. On the other hand, we show that this version of the problem is fixed-parameter tractable for planar graphs, using problem-specific techniques; see Section 8.3.

### 4.4 Geometric Graphs

In some of the applications, the environment can be naturally modeled by the infinite geometric graph defined by Euclidean space, where vertices correspond to points and edges connect two vertices that are within a fixed distance of each other, say 1. In this case, we develop efficient algorithms in a very general setting in Section 8.4, even though the graph is infinite:

**Theorem 4.2.** *If $\mathcal{G}$ is any regular graph property, then given rational starting coordinates for $k$ total agents (including facility and obnoxious agents) in Euclidean $d$-space, we can find a solution to the movement problem up to additive error $\varepsilon > 0$ using $f(k,d) \cdot n^{O(1)}(\lg D + \lg(1/\varepsilon))$ time, where $D$ is the maximum distance between any two starting coordinates.*

The main tool for proving this theorem is Canny's Roadmap Algorithm for motion planning in Euclidean space [5], which lets us manipulate bounded-size semi-algebraic sets; see Section 8.4.

### 4.5 Improving CONNECTIVITY with Fast Subset Convolution

Finally, we optimize one particularly practical problem, CONNECTIVITY: moving the agents so that they form a connected subgraph. Our general characterization implies that this problem is fixed-parameter tractable. Using the recent algorithm of Björklund et al. [3] for fast subset convolution in the min-sum semiring, in Section 8.5 we design a more efficient algorithm for this problem: the exponential factor of the running time is only $O(2^k)$.

In summary, our results form a systematic study of the movement problem, using powerful tools to classify the complexity of the different variants. Our algorithms are general, so may not be optimal for any specific version of the problem, but they nonetheless characterize which problems are tractable, and lead the way for future investigation into more efficient algorithms for practical special cases.

## 5 Model and Definitions

In this section, we make precise the model described in Section 1 and introduce some additional notation.

**Definition 5.1.** *We fix three finite sets of colors: $C_m$ (main colors), $C_f$ (facility colors), $C_o$ (obnoxious colors).*

Pebbles with main colors will represent the client agents: the target pattern has to contain all such pebbles. Pebbles with facility colors are "good": having more than the prescribed number of such pebbles is still an acceptable solution. Conversely, pebbles with obnoxious colors are "bad": removing such a pebble from a target pattern does not make a solution invalid.

**Definition 5.2.** *A* multicolored graph *is a graph with a multiset of colored pebbles assigned to each vertex (a vertex can be assigned multiple pebbles with the same color).*

We extend the notions of vertex removal, edge removal, edge addition, and induced subgraphs to multicolored graphs the obvious way, i.e., the set pebbles at the vertices (remaining in the graph) is unchanged.

**Definition 5.3.**
- We denote by $n_G(c, v)$ the number of pebbles with color $c$ at vertex $v$ in $G$.

- A *multicolored graph property* is a (possibly infinite) recursively enumerable set $\mathcal{G}$ of multicolored graphs.

- A graph property $\mathcal{G}$ is *regular* if for every fixed $k, \ell$ there is only a finite number of graphs in $\mathcal{G}$ with at most $\ell$ vertices and at most $k$ main pebbles and there is an algorithm that, given $k$ and $\ell$, enumerates these graphs. (Note that the number of facility and obnoxious pebbles is not bounded here.)

- A graph property $\mathcal{G}$ is *hereditary* if, for every $G \in \mathcal{G}$, every induced multicolored subgraph of $G$ is also in $\mathcal{G}$.

- A graph property $\mathcal{G}$ is *closed under edge addition* if whenever $G$ is in $\mathcal{G}$ and $G'$ is the multicolored graph obtained from $G$ by connecting two nonadjacent vertices, then $G'$ is also in $\mathcal{G}$.

- A graph $G \in \mathcal{G}$ is *edge-deletion minimal* if there is no multicolored graph $G' \in \mathcal{G}$ that can obtained from $G$ by edge deletions.

**Definition 5.4.** *Let $G_1$ and $G_2$ be two multicolored graphs whose underlying graphs are isomorphic. $G_2$ dominates $G_1$ if there is an isomorphism $\phi : V(G_1) \to V(G_2)$ such that, for every $v \in V(G_1)$,*

1. *for every $c \in C_m$, vertices $v$ and $\phi(v)$ have the same number of pebbles with color $c$;*

2. *for every $c \in C_f$, vertex $\phi(v)$ has at least as many pebbles with color $c$ as $v$; and*

3. *for every $c \in C_o$, vertex $\phi(v)$ has at most as many pebbles with color $c$ as vertex $v$.*

**Definition 5.5.** *For every set $\mathcal{G}$ of multicolored graphs, the* movement problem *has the following inputs:*

1. *a multicolored graph $G(V, E)$, $P$ is the set of pebbles, $k$ is the number of main pebbles;*

2. *a movement cost function $c_p : V \to Z^+$ for each pebble $p \in P$;*

3. *integer $\ell$, the maximum solution size; and*

4. *integer $C$, the maximum cost.*

*The task is to find a movement plan $m : P \to V$ such that*

1. *the total cost $\sum_{p \in P} c_p(m(p))$ of the moves is at most $C$; and*

2. *after the movements, there is a set $S$ of at most $\ell$ vertices such that $S$ contains all the main pebbles and the multicolored graph $G[S]$ dominates some graph in $\mathcal{G}$.*

By using different movement cost functions, we can express various goals:

1. if $c_p(v)$ is the distance of $p$ from $v$, then we have to minimize the sum of movements,

2. if $c_p(v) = 0$ if $v$ is at distance at most $d$ from $p$ and $\infty$ otherwise, then we have to find a solution where $p$ moves at most $d$ steps,

3. if $c_p(v) = 0$ if $v$ is the initial location of $p$ and $c_p(v) = 1$ for every other vertex, then we have to minimize the number of pebbles that move.

Of course, we can express combinations of these goals or the distance can be measured on different graphs for the different pebbles, etc. The formulation is very flexible.

# 6  Main Algorithm

In this section, we present our main algorithm, i.e., the proof of Theorem 3.1. The algorithm is based on enumerating minimal configurations, nontrivially using the color-coding technique to narrow the possibilities. Then it finds the best possible location in the graph to realize each minimal configuration. By assumption, the minimal configurations have bounded treewidth and, as the following classical result shows, finding such subgraphs is FPT (see also [28]):

**Theorem 6.1** ([1]). *Let $F$ be an undirected graph on $k$ vertices with treewidth $t$. Let $G$ be an undirected graph with $n$ vertices. A subgraph of $G$ isomorphic to $F$, if one exists, can be found in time $2^{O(k)} \cdot n^{O(t)}$.*

However, we need a weighted version of this result in order to express the movement costs. A *subgraph embedding* of $F$ in $G$ is a mapping $\phi : V(F) \to V(G)$ such that if $u, v \in V(F)$ are adjacent in $F$, then $\phi(u)$ and $\phi(v)$ are adjacent in $G$. Let $c : V(F) \times V(G) \to \mathbb{Z}^+$ be a cost function that determines the cost of mapping a vertex of $F$ to a vertex of $G$. If $\phi$ is a subgraph embedding of $F$ into $G$, then we define the cost of $\phi$ to be $c(\phi) := \sum_{v \in V(F)} c(v, \phi(v))$. Note that $c(v_1, u)$ is not necessarily equal to $c(v_2, u)$, thus the cost of mapping two different vertices $v_1$ and $v_2$ of $F$ to a particular vertex $u$ of $G$ can have different costs. By extending the techniques of Theorem 6.1, we can find a subgraph embedding of minimum cost.

**Theorem 6.2.** *Let $F$ be an undirected graph on $k$ vertices with treewidth $t$. Let $G$ be an undirected graph with $n$ vertices and let $c : V(F) \times V(G) \to \mathbb{Z}^+$ be a cost function of mapping a vertex of $F$ to a vertex of $G$. If $F$ is a subgraph of $G$, then it is possible to find in time $2^{O(k)} \cdot n^{O(t)}$ a subgraph embedding $\phi$ that minimizes $c(\phi)$.*

Similarly to Theorem 6.1, the proof of Theorem 6.2 is based on dynamic programming on the tree decomposition of $F$. However, here we have to maintain minimum-cost solutions instead of feasibility. This modification of the proof is quite straightforward, but as the proof in [1] is rather sketchy, we give a full proof in the Appendix for completeness.

*Proof (of Theorem 3.1).* In the solution, the set $S \subseteq V(G)$ has to induce a graph $F$ that dominates some graph $F' \in \mathcal{G}$. The graph $F'$ has a subgraph $F_0$ that is an edge-deletion minimal graph of $\mathcal{G}$. Let $\mathcal{G}_{k,\ell}$ be the set of edge-deletion minimal graphs in $\mathcal{G}$ with at most $\ell$ vertices and exactly $k$ main pebbles. Since $\mathcal{G}$ is regular, $\mathcal{G}_{k,\ell}$ is finite and we can enumerate the graphs in $\mathcal{G}_{k,\ell}$ in time depending only on $k$ and $\ell$. Let us denote by $D_{k,\ell}$ the maximum number of pebbles in a graph of $\mathcal{G}_{k,\ell}$. For each $F_0 \in \mathcal{G}_{k,\ell}$, we test whether there is a solution with this particular $F_0$.

For a given $F_0$, we proceed as follows. Let $v_1, \ldots, v_{\ell_0}$ be the vertices of $F_0$ (note that $\ell_0 \leq \ell$). A solution consists of two parts: a subgraph embedding of $F_0$ into $G$ and a way of moving the pebbles. Formally, a solution is a pair $(\phi, m)$ where $\phi : V(F_0) \to V(G)$ is a subgraph embedding of $F_0$ in $G$ and $m : P \to V(G)$ describes how the pebbles are moved. The objective is to minimize the total cost $\sum_{p \in P} c_p(m(p))$ of the movements. For a given embedding $\phi$, there might be several possible movement plans $m$ such that $(\phi, m)$ is a solution, i.e., the vertices in $\phi(V(F_0))$ have the appropriate pebbles after the movements of $m$. Thus for each embedding $\phi$, there is a minimal cost $c_{\min}(\phi)$ of a movement plan that forms a solution together with $\phi$. Therefore, we have to find an embedding $\phi$ such that $c_{\min}(\phi)$ is minimal. The main idea of the proof is to try to express this cost $c_{\min}(\phi)$ as a linear function of the mapping, i.e, as $c_{\min}(\phi) = \sum_{v \in V(F)} c(v, \phi(v))$ for some appropriate function $c$. If we can do that, then Theorem 6.2 can be invoked to find the embedding $\phi$ with the smallest $c_{\min}(\phi)$. However, it seems unlikely that the cost of the best way of moving the pebbles can be expressed as a simple linear function of the embedding. If the embedding $\phi$ is fixed, finding the best movement plan involves making non-independent decisions about which pebble goes where, hence $c_{\min}(\phi)$ seems to be a very nonlinear function. What we do instead is to make some guesses about the internal structure of the solution, and construct a linear cost function that is correct for solutions with such structure.

Let $L$ be a random labeling that assigns labels from $\{1, \ldots, \ell\}$ to the main and facility pebbles, and labels from $\{0, 1, \ldots, \ell\}$ to the vertices of $G$. We say that a solution $(\phi, m)$ is *$L$-good* if

(R1)  for $1 \leq i \leq \ell_0$, vertex $\phi(v_i)$ has label $i$;

(R2)  if $m(p) = \phi(v_i)$ for a main pebble $p$, then $p$ has label $i$;

(R3) there are at least $n_{F_0}(v_i, c)$ facility pebbles $p$ having label $i$ such that $m(p) = \phi(v_i)$; and

(R4) for every $1 \le i \le \ell_0$, if $p$ is an obnoxious pebble initially located at $\phi(v_i)$, then either $m(p) = \phi(v_j)$ for some $1 \le j \le \ell_0$, or $m(p)$ is a vertex with label 0.

We show that restricting our attention to $L$-good solutions is not a serious restriction, as we can bound from below the probability that a fixed solution is $L$-good with respect to a random labeling:

**Lemma 6.3.** *Let $(\phi, m)$ be an optimum solution that moves the minimum number of pebbles. Solution $(\phi, m)$ is $L$-good with respect to a random labeling $L$ with positive probability depending only on $k$, $\ell$, and $|C_0|$.*

*Proof.* Requirement (R1) holds with probability $(\ell+1)^{-\ell_0}$. There are at most $D_{k,\ell}$ facility pebbles in $F_0$, thus there are no reason to move more than $D_{k,\ell}$ facility pebbles. Requirements (R2) and (R3) prescribe specific labels on at most $D_{k,\ell}$ pebbles. Finally, observe that if $p_1, p_2$ are two pebbles with color $c \in C_o$ initially located at $\phi(v_i)$ and $m(p_1), m(p_2) \notin \phi(V(F_0))$, then it can be assumed that $m(p_1) = m(p_2)$ (here we use that the movement cost functions of $p_1$ and $p_2$ are assumed to be the same, hence if both pebbles are moved outside $\phi(V(F_0))$, then we can move them to the same vertex). Thus it can be assumed that there are at most $\ell_0|C_o|$ vertices outside $\phi(V(F_0))$ where obnoxious pebbles are moved to, i.e., (R4) requires label 0 on at most $\ell_0|C_o|$ vertices. Since the requirements are independent, a random labeling satisfies all of them with positive probability. □

Let us fix a subgraph embedding $\phi : V(F_0) \to V(G)$ and let us intuitively discuss what is the cost of moving the pebbles in an $L$-good solution $(\phi, m)$. The cost comes from three parts: moving the main, the facility, and the obnoxious pebbles.

- **Main pebbles.** As $\phi$ is $L$-good, every main pebble with label $i$ should go to $\phi(v_i)$. This means that $\phi$ determines the cost of moving the main pebbles.

- **Facility pebbles.** For every color $c$, we have to ensure that at least $n_{F_0}(v_i, c)$ facility pebbles with color $c$ and label $i$ are moved to $\phi(v_i)$ (including the possibility that some of them were already there and stay there). It is clear that we can always do this the cheapest possible way, i.e., by selecting those $n_{F_0}(v_i, c)$ pebbles with color $c$ and label $i$ whose cost of moving to $\phi(v_i)$ is minimum possible. In particular, no conflict arises between moving vertices to $\phi(v_i)$ and to $\phi(v_j)$, as they involve only vertices with label $i$ and $j$, respectively.

- **Obnoxious pebbles.** We have to ensure that at most $n_{F_0}(v_i, c)$ obnoxious pebbles of color $c$ remains at $\phi(v_i)$ after the movement. The rest should be moved somewhere else, preferably not to any other $\phi(v_j)$. If the cheapest way of moving an obnoxious pebble away from $\phi(v_i)$ is outside $\phi(V(F_0))$, then we should definitely move the pebbles there. However, it could be possible that the cheapest place is inside $\phi(V(F_0))$ and therefore in an optimum solution we cannot avoid moving an obnoxious pebble from $\phi(v_i)$ to some $\phi(v_j)$. To account for these movements, we guess the exact way the obnoxious pebbles move inside $\phi(V(F_0))$. For every $1 \le i, j \le \ell_0$ and $c \in C_o$, denote by $e_{c,i,j}$ the number of pebbles with color $c$ that is moved from $\phi(v_i)$ to $\phi(v_j)$ (in particular, $e_{c,i,i}$ is the number of pebbles with color $c$ that stay at $\phi(v_i)$). The tuple $E$ of these $\ell_0^2|C_o|$ numbers will be called the *scheme* of the solution. Observe that $e_{c,i,j} \le D_{k,\ell}$, thus there are $D_{k,\ell}^{\ell_0^2|C_o|}$ possible schemes, which is a constant depending only on $k$, $\ell$, and the property $\mathcal{G}$. We say that a scheme is *correct* if for every $c \in C_o$ and $1 \le j \le \ell_0$, the sum $\sum_{i=1}^{\ell_0} e_{c,i,j} \le n_{F_0}(c, v_j)$, i.e., the scheme does not move more obnoxious pebbles to a vertex $v_j$ than it is allowed there. It is clear that the scheme of a solution is always correct. The embedding $\phi$ and the scheme of the solution determines the way the obnoxious pebbles are moved: $e_{c,i,j}$ tells us how many pebbles of color $c$ have to be moved from $\phi(v_i)$ to $\phi(v_j)$ and the remaining pebbles should be moved to the closest vertex with label 0. Furthermore, if an obnoxious pebble is outside $\phi(V(F_0))$ initially, then there is no reason to move it.

As mentioned earlier, we cannot make $c_{\min}(\phi)$ a linear function. However, we can make it a linear function for $L$-good embeddings with respect to a particular labeling $L$ and scheme $E$, in the following sense:

**Claim 6.4.** *Let $L$ be a labeling and $E$ be a correct scheme. It is possible to define an embedding function $c(v_i, u)$ with the following properties:*

(P1) If $(\phi, m)$ is an L-good solution with the scheme $E$, then $c(\phi)$ is at most the cost of $(\phi, m)$.

(P2) If $F_0$ has an embedding $\phi'$ into $G$, then there is a (not necessarily L-good) solution $(\phi', m)$ with cost at most $c(\phi')$.

*Proof.* The embedding cost $c(v_i, u)$ of mapping $v_i \in V(F_0)$ to $u \in V(G)$ is defined to be the sum of 3 terms:

1. The total cost $c_1(v_i, u)$ of moving all the main pebbles with label $i$ to $u$.

2. The total cost $c_2(v_i, u)$ of moving, for every color $c \in C_f$, $n_{F_0}(v_i, c)$ facility pebbles with label $i$ to $u$. If there are more than $n_{F_0}(v_i, c)$ such facility pebbles, then we move those whose movement cost to $u$ is minimal. If there are less than $n_{F_0}(v_i, c)$ such facility pebbles, then we make the cost infinite.

3. The total cost $c_3(v_i, u)$ of moving away the obnoxious pebbles from $u$, according to the scheme. For a color $c \in C_o$, let $t(c, u, j)$ be the minimum cost of moving a pebble with color $c$ from $u$ to a vertex with label $j$. The total cost of removing the required number of obnoxious pebbles from $u$ is

$$c_3(v_i, u) = \sum_{c \in C_o} \left( \sum_{j=1}^{\ell_0} e_{c,i,j} \cdot t(c, u, j) + \left( n_G(u, c) - n_{F_0}(v_i, c) - \sum_{j=1}^{\ell_0} e_{c,i,j} \right) t(c, u, 0) \right).$$

Additionally, if $u$ is a vertex with label *different* from $i$, then we make the cost $c(v_i, u)$ infinite. It is straightforward to see that (P1) holds: the three components of $c(\phi)$ are covered by the cost of moving the pebbles. The cost of moving a main or facility pebble contributes to the embedding cost of the vertex where it arrives, while the cost of moving an obnoxious vertex contributes to the embedding cost of the vertex where it was initially.

To see that (P2) holds, let $\phi'$ be an embedding of $F_0$ into $G$ with cost $c(\phi')$. We construct a solution with cost at most $c(\phi')$ where $\phi'(V(F_0))$ induces a multicolored graph that has a subgraph dominating $F_0$. First we move every main pebble with label $i$ to vertex $\phi'(v_i)$. The cost of this is covered by the first component of the cost function. Next for every $c \in C_f$ and $1 \leq i \leq \ell_0$, we move $n_{F_0}(v_i, c)$ pebbles with color $c$ and label $i$ to $\phi'(v_i)$. Note that there are at least $n_{F_0}(v_i, c)$ such pebbles: otherwise the cost would be infinite by our definition. If there are more than $n_{F_0}(v_i, c)$ such pebbles, then we move those pebbles whose cost of moving to $v_i$ is minimal. The total cost of this is covered by the second component of $c(\phi')$.

Finally, we move the obnoxious pebbles according to the scheme. For each $c \in C_o$ and $1 \leq i \leq \ell_0$, we move $e_{c,i,j}$ of the pebbles at $\phi'(v_i)$ to a vertex with label $j$. This incurs a total cost of $\sum_{c \in C_o} \sum_{j=1}^{\ell_0} e_{c,i,j} \cdot t(c, \phi'(v_i), j)$ when moving the pebbles initially located at $\phi'(v_i)$. After that, for each $c \in C_0$, we have to move from $\phi'(v_i)$ some of the remaining pebbles with color $c$ to ensure that only $n_{F_0}(v_i, c)$ such pebbles with color $c$ remain at $\phi'(v_i)$. We move these pebbles to the closest vertex having label 0, thus the total cost of these movements is $\sum_{c \in C_o} (n_G(\phi'(v_i), c) - n_{F_0}(v_i, c) - \sum_{j=1}^{\ell_0} e_{c,i,j}) t(c, \phi'(v_i), 0))$. Clearly, the third component of the cost covers the cost of these moves. As vertex $\phi'(v_i)$ is a vertex with label $i$ (otherwise $c(v_i, u)$ would be infinite), there are at most $\sum_{j=1}^{\ell_0} e_{c,j,i}$ pebbles with color $c$ that goes to $\phi'(v_i)$, which is at most $n_{F_0}(v_i, v)$ since the scheme is correct. □

Note that the solution $(\phi', m)$ is not necessarily $L$-good and does not necessarily respect the scheme $E$. However, we eventually do not care about internal properties of the solution other than the cost. Observe that the only reason why $(\phi', m)$ is not $L$-good is that obnoxious pebbles can be moved to a vertex of label $i$ different from $\phi(v_i)$, but this just means that fewer obnoxious pebbles are moved to $\phi(v_i)$ than expected.

In summary, the algorithm performs the following steps:

1. Try every $F_0 \in \mathcal{G}_{k,\ell}$ and try every correct scheme $E$.

2. Take a random labeling $L$ of the pebbles and the vertices.

3. Based on $F_0$, the scheme $E$, and the labeling, construct the embedding cost function $c(v_i, u)$ defined by Claim 6.4.

4. Using Theorem 6.2, find the minimum cost subgraph embedding $\phi$ with this cost function.

5. Construct the solution $(\phi, m)$ defined by (P2) of Claim 6.4.

We claim that the above algorithm finds an optimum solution with positive probability depending only on $k$, $\ell$, $\mathcal{G}$, thus by repeating the algorithm $f(k, \ell, \mathcal{G})$ times, the error probability can be made arbitrarily small. (The algorithm can be derandomized by using $k$-perfect families of hash functions instead of the random labeling [1], [16, Section 13.3]; we omit the details.) Let $(\phi, m)$ be an optimum solution with cost OPT and let $F_0 \in \mathcal{G}_{k,\ell}$ be the edge-deletion minimal graph of $\mathcal{G}$ corresponding to the solution. At some point, the algorithm considers this particular $F_0$ and the scheme $E$ of this solution. In Step 2, with constant probability, this particular solution $(\phi, m)$ is $L$-good (as discussed above). Thus, by (P1) of Claim 6.4, the embedding cost of $\phi$ is at most OPT. This means that in Step 5, we find an embedding with cost at most OPT, and in Step 6, by (P2) of Claim 6.4, we find a solution with cost at most OPT.

The number of possibilities tried in Step 1 is a constant depending only on $k$ and $\ell$. The application of Theorem 6.2 in Step 5 takes time $f(\ell)n^{O(w)}$, where $w$ is the maximum treewidth of an edge-deletion minimal graph in $\mathcal{G}$, which is a constant depending only on $\mathcal{G}$ (and not on $k$ and $\ell$). Every other step is polynomial. Thus the running time is $f(k, \ell) \cdot n^{O(1)}$ for a fixed $\mathcal{G}$. □

# 7 Main Hardness Proof

In this section, we prove the main hardness result, Theorem 3.2. The proof uses a result on the structural complexity of constraint satisfaction problems. An instance of a *constraint satisfaction problem* is a triple $(V, D, C)$, where

1. $V$ is a set of variables,

2. $D$ is a domain of values,

3. $C$ is a set of constraints, $\{c_1, c_2, \ldots, c_q\}$. Each constraint $c_i \in C$ is a pair $\langle s_i, R_i \rangle$, where

    (a) $s_i$ is a tuple of variables of length $m_i$, called the *constraint scope*, and

    (b) $R_i$ is an $m_i$-ary relation over $D$, called the *constraint relation*.

For each constraint $\langle s_i, R_i \rangle$ the tuples of $R_i$ indicate the allowed combinations of simultaneous values for the variables in $s_i$. The length $m_i$ of the tuple $s_i$ is called the *arity* of the constraint. A *solution* to a constraint satisfaction problem instance is a function $f$ from the set of variables $V$ to the domain of values $D$ such that for each constraint $\langle s_i, R_i \rangle$ with $s_i = \langle v_{i_1}, v_{i_2}, \ldots, v_{i_{m_i}} \rangle$, the tuple $\langle f(v_{i_1}), f(v_{i_2}), \ldots, f(v_{i_{m_i}}) \rangle$ is a member of $R_i$. We say that an instance is *binary* if each constraint relation is binary, i.e., $m_i = 2$ for each constraint.

The *primal graph* of a binary constraint satisfaction instance is a graph where the vertices are the variables and the edges are the constraints. Given a class $\mathcal{G}$ of graphs, we denote by BINARY-CSP($\mathcal{G}$) the restriction of binary CSP to instances whose primal graph is in $\mathcal{G}$. It is well-known that if $\mathcal{G}$ has bounded treewidth, then BINARY-CSP($\mathcal{G}$) is polynomial-time solvable [17]. The converse is also true:

**Theorem 7.1** (Grohe [19]). *If $\mathcal{G}$ is a recursively enumerable class of graphs with unbounded treewidth, then BINARY-CSP($\mathcal{G}$) parameterized by the number of variables is W[1]-hard.*

We prove Theorem 3.2 using Theorem 7.1.

*Proof (of Theorem 3.2).* Let $\mathcal{G}_0$ contain the underlying graphs (i.e., disregarding pebbles) of all the multi-colored graphs $G_w$ ($w \geq 1$) defined in the statement of the theorem. It is easy to see that $\mathcal{G}_0$ is recursively enumerable. By assumption, $\mathcal{G}_0$ has unbounded treewidth, hence BINARY-CSP($\mathcal{G}_0$) is W[1]-hard by Theorem 7.1. We present a reduction from BINARY-CSP($\mathcal{G}_0$) to the movement problem.

Consider an instance of BINARY-CSP($\mathcal{G}_0$) with primal graph $F_0 \in \mathcal{G}_0$. Let $x_1, \ldots, x_\ell$ be the variables of $F_0$ and let $\{1, 2, \ldots, m\}$ be the domain $D$. By the definition of $\mathcal{G}_0$, there is an edge-deletion minimal multicolored graph $F \in \mathcal{G}$ whose underlying graph is $F_0$; let $u_1, \ldots, u_\ell$ be the vertices of $F$. (Since $\mathcal{G}$ is recursively enumerable, such an $F$ can be found in time depending only on the size of $F_0$.) Let $p_1, \ldots, p_k$ be the pebbles in $F$ (by assumption, all these pebbles are main pebbles); assume that pebble $p_i$ is on vertex $u_{t_i}$. We construct a graph $G$ as follows. Graph $G$ has $|D|\ell + k$ vertices: $v_{i,j}$ for $1 \leq i \leq \ell$ and $1 \leq j \leq m$ and $q_i$ for $1 \leq i \leq k$. If pebble $p_i$ is on vertex $u_{t_i}$ in the graph $F$, then vertex $q_i$ of $G$ is connected to every vertex $v_{t_i,j}$ with $1 \leq j \leq m$. For each binary constraint $\langle(x_{i_1}, x_{i_2}), R\rangle$, we add edges as follows: for every pair $(d_{i_1}, d_{i_2}) \in R$, we connect the vertices $v_{i_1,d_{i_1}}$ and $v_{i_2,d_{i_2}}$. To complete the description of the instance, we put the pebble $p_i$ on $q_i$ for every $i$, $1 \leq i \leq k$.

We claim that the CSP instance has a solution if and only if there is a solution for the movement problem on at most $\ell$ vertices such that each pebble moves at most one step. Assume that $f : V \to D$ is a solution for the CSP instance. For every $1 \leq i \leq k$, we move pebble $p_i$ from $q_i$ to $v_{t_i, f(x_{t_i})}$. We show that the pebbles induce a multicolored graph isomorphic to $F$. First, observe that all the pebbles are on the $\ell$ vertices $v_{1,f(x_1)}$, ..., $v_{\ell,f(x_\ell)}$ and $v_{i,f(x_i)}$ has the same number and types of pebbles as $u_i$ in $F$. If $u_{i_1}$ and $u_{i_2}$ are connected in $F$, then there is a binary constraint $\langle(x_{i_1}, x_{i_2}), R\rangle$ in the CSP instance. As $f$ is a solution, we have $(f(x_{i_1}), f(x_{i_2})) \in R$ and hence there is an edge connecting $v_{i_1,f(x_{i_1})}$ and $v_{i_2,f(x_{i_2})}$. On the other hand, if $u_{i_1}$ and $u_{i_2}$ are not adjacent in the primal graph, then there is no edge connecting any $v_{i_1,j_1}$ with any $v_{i_2,j_2}$.

Conversely, we show that if there is a solution for the movement problem, then there is a solution $f$ for the CSP instance. Each pebble can move only one step, thus pebble $p_i$ either stays at $q_i$ or goes to $v_{t_i,j}$ for some $1 \leq j \leq m$. Therefore, at least $\ell$ vertices contain pebbles after the moves. We assumed that the solution is on at most $\ell$ vertices, hence for every $1 \leq i \leq \ell$, there is at most one $1 \leq j \leq m$ such that there is a pebble on $v_{i,j}$. Define $f(x_i)$ to be this value $j$, or define $f(x_i)$ arbitrarily if there is no such $j$ (i.e., all the pebbles adjacent to $v_{i,1}$, ..., $v_{i,|D|}$ remained at their initial location). Observe that if two pebbles in $G$ are collocated after the moves, then they are collocated in $F$ as well. Therefore, the pebbles occupy at least $|V(F)| = \ell$ vertices. However, we know that in the solution the pebbles occupy exactly $\ell$ vertices, which is only possible if whenever two pebbles are collocated in $F$, then they are collocated in the solution. If pebbles $p_{i_1}$ and $p_{i_2}$ are not neighbors in $F$, then they cannot be neighbors after the moves, since in that case there is no edge connecting any $v_{t_{i_1},j_1}$ with any $v_{t_{i_2},j_2}$. Thus the graph induced by the pebbles is a subgraph of $F$. However, as $F$ is an edge-deletion minimal graph of $\mathcal{G}$, this is only possible if the pebbles induce the graph $F$ itself. This means that for every edge $v_{i_1}v_{i_2}$ of $F$, vertices $v_{i_1,f(x_{i_1})}$ and $v_{i_2,f(x_{i_2})}$ have to be neighbors. By the way the graph was defined, this is only possible if $(f(x_{i_1}), f(x_{i_2})) \in R$, that is, $f$ satisfies the constraint. □

## 8 Further Techniques

In this section, we prove the various results described in Section 4, as well as the hardness result for hereditary problems described in Sections 2 and 3.

### 8.1 Planar Graphs and $H$-Minor-Free Graphs

To prove Theorem 4.1, we need a version of Theorem 6.2 that finds a minimum cost *induced subgraph embedding,* i.e., a mapping $\phi : V(F) \to V(G)$ such that $\phi(v_1)\phi(v_2)$ is an edge of $G$ if and only if $v_1v_2$ is an edge of $F$. First, the minimum cost induced subgraph embedding can be found in linear time if $G$ has bounded treewidth:

**Theorem 8.1.** *Let $F$ be an undirected graph on $k$ vertices, let $G$ be an undirected graph with treewidth $w$, and let $c : V(F) \times V(G) \to \mathbb{Z}^+$ be a cost function of mapping a vertex of $F$ to a vertex of $G$. It is possible to find in time $f(k,w) \cdot n$ an induced subgraph embedding $\phi$ of $F$ into $G$ that minimizes $c(\phi)$ (if such an embedding exists).*

The proof of Theorem 8.1 uses the standard algorithmic techniques of bounded treewidth graphs (see e.g., [14] for a similar result). We omit the details. To generalize Theorem 8.1 to the case when $G$ is planar, or more generally, $H$-minor-free, we need the following result:

**Theorem 8.2.** [10] *For a fixed graph $H$, there is a constant $c_H$ such that, for any integer $k \geq 1$ and for every $H$-minor-free graph $G$, the vertices of $G$ can be partitioned into $k+1$ sets such that any $k$ of the sets induce a graph of treewidth at most $c_H k$. Furthermore, such a partition can be found in polynomial time.*

**Theorem 8.3.** *Let $F$ be an undirected graph on $k$ vertices, let $G$ be an undirected graph, and let $c : V(F) \times V(G) \to \mathbb{Z}^+$ be a cost function of mapping a vertex of $F$ to a vertex of $G$. For every fixed graph $H$,* *if $G$ is $H$-minor-free, then it is possible to find in time $f(k) \cdot n^{O(1)}$ an induced subgraph embedding $\phi$ of $F$ into $G$ that minimizes $c(\phi)$ (if such an embedding exists).*

*Proof.* Let $k := |V(F)|$ and let $V_1, \ldots, V_{k+1}$ be the partition of $V(G)$ obtained by Theorem 8.2. If $\phi$ is minimum cost embedding, then by the pigeon hole principle, there is a $1 \leq i \leq k+1$ such that $\phi$ does not map vertices to $V_i$. This means that the minimum cost embedding of $F$ into $G \setminus V_i$ has the same cost as $\phi$. For every $1 \leq i \leq k+1$, we find the minimum cost embedding $\phi_i$ from $F$ to $G \setminus V_i$; clearly, the $\phi_i$ with minimum cost gives an optimum solution. Because the treewidth of $G \setminus V_i$ is at most $c_H k$, by Theorem 8.1, embedding $\phi_i$ can be found in time $f(k, c_H k)n$. This means that for a fixed $H$, the whole algorithm needs $f'(k)n^{O(1)}$ time for some function $f'$. □

With these tools in hand, the proof of Theorem 4.1 is the same as the proof of Theorem 3.1, but with the following differences:

1. $\mathcal{G}_{k,\ell}$ contains every graph with $k$ main pebbles and at most $\ell$ vertices, not only the edge-deletion minimal ones.

2. For a given $F_0 \in \mathcal{G}_{k,\ell}$, we find a minimum cost *induced* subgraph embedding using Theorem 8.3, instead of a minimum cost subgraph embedding using Theorem 6.2.

## 8.2 Bidimensionality

Theorem 4.1 shows that the DISPERSION problem defined in Section 2.1 is FPT for planar graphs, parameterized by the number $k$ of pebbles. Note that Theorem 4.1 holds for arbitrary movement cost functions. In the special case where each pebble is allowed to move only one step in the planar graph, called SINGLE-MOVE DISPERSION, we show how to obtain a simpler and more efficient algorithm using bidmensionality theory.

We argue as follows. If a vertex does not have a pebble in its closed neighborhood, then it is irrelevant to the problem and can be deleted without changing the answer. Next we establish a parameter-treewidth bound, as in bidimensionality theory [9]: if the treewidth is sufficiently large, then the parameter must be larger than $s$, so we can simply answer "no". To prove this relation we use the grid-minor theorem of [30]: there is a universal constant $c$ such that if a planar graph has treewidth at least $c \cdot s$, then it has a $s \times s$ grid as a minor. Now suppose that the treewidth of the graph is at least $4c\lceil\sqrt{k}\rceil$, which implies that the graph has a $4\lceil\sqrt{k}\rceil \times 4\lceil\sqrt{k}\rceil$ grid minor. If we consider just the edge contractions that lead to this minor, ignoring the edge deletions, we obtain a partially triangulated grid as in bidimensionality theory [9]. Let $S$ be the set of grid vertices interior to the grid (excluding the boundary) and having row and column numbers divisible by 3. In this way, we find a set of at least $k+1$ vertices such that any two vertices are at distance more than 2 from each other. There is a pebble in the closed neighborhood of each vertex $v \in S$ (otherwise $v$ would be irrelevant). But the closed neighborhoods are disjoint, implying that that there are at least $k+1$ pebbles, a contradiction.

Thus we can assume that the treewidth of the graph is at most $4c\lceil\sqrt{k}\rceil$. In this case, we can solve the problem with a simple application of Courcelle's Theorem [8]. Sentences in the *Extended Monadic Second Order Logic of Graphs* (EMSO) contain quantifiers, logical connectives ($\neg$, $\vee$, and $\wedge$), vertex variables, edge variables, vertex set variables, edge set variables, and the following binary relations: $\in$, $=$, $\text{inc}(e,v)$ (edge variable $e$ is incident to vertex variable $v$), and $\text{adj}(u,v)$ (vertex variables $u$, $v$ are neighbors). Furthermore, the language can contain arbitrary unary predicates on the vertices and edges. If a graph property can be expressed in EMSO then, for every fixed $w$, the problem can be solved in linear time on graphs with treewidth at most $w$ [8]. It is easy to verify that the following formula expresses that the SINGLE-MOVE DISPERSION with $k$ pebbles has a solution. The predicate $P(v)$ expresses that there is a pebble at vertex $v$ in the initial graph.

$$\exists x_1, \ldots, x_k, y_1, \ldots, y_k : \bigwedge_{1 \leq i \leq k} (P(x_i) \wedge (\text{adj}(x_i, y_i) \vee x_i = y_i))$$
$$\wedge \bigwedge_{1 \leq i < j \leq k} (x_i \neq x_j \wedge y_i \neq y_j \wedge \neg\text{adj}(y_i, y_j))$$

Interestingly, one can give a single EMSO formula for SINGLE-MOVE DISPERSION that does not depend on the number $k$ of pebbles (we can describe by a set $E$ of edges the moves in the solution; we omit the details). Either way, we get that for every fixed $k$, the problem can be solved in linear time on graphs with treewidth at most $4c\lceil\sqrt{k}\rceil$, completing the algorithm.

**Theorem 8.4.** SINGLE-MOVE DISPERSION *can be solved in* $f(k) \cdot n^{O(1)}$ *time.*

Courcelle's Theorem gives an easy way of showing that a certain problem is FPT, but it cannot be used to optimize the exact running time, i.e., the function $f(k)$. We believe we can obtain a better running time via a direct dynamic programming algorithm for SINGLE-MOVE DISPERSION on bounded-treewidth graphs. We do not give an explicit description of this, as the techniques are standard and the details are somewhat tedious. The main observation that if the tree decomposition has width $w$, then the dynamic programming has to consider $2^{O(w)}$ states describing which vertices of the bag have pebbles on it in the solution and whether any pebble initially located in the bag was moved to a bag lower in the decomposition. This would take more work than applying Courcelle's Theorem, but results in an algorithm where the function $f(k)$ in the running time is only $2^{O(\sqrt{k})}$, which is subexponential in $k$. It remains an interesting question for further work to determine whether other variants of the problem (such having an unbounded number of pebbles but parameterizing by the total movement) admit subexponential-time FPT algorithms based on bidimensionality.

## 8.3 Planar STEINER CONNECTIVITY

We have seen in Section 3.1 that STEINER CONNECTIVITY (connect the red pebbles using the blue pebbles) is FPT parameterized by the total number of pebbles (red and blue). On the other hand, if the parameter is only the number of blue pebbles, then the problem is W[2]-hard:

**Theorem 8.5.** STEINER CONNECTIVITY *is W[2]-hard parameterized by the number of blue pebbles, even in the special case when each blue pebble is allowed to move at most one step in the graph (and the red pebbles are stationary).*

*Proof.* The proof is similar to the proof Theorem 3.5. We present a reduction from MINIMUM DOMINATING SET: given a graph $G$ with $n$ vertices and an integer $k$, we construct an instance of STEINER CONNECTIVITY with $n$ red pebbles and $k$ blue pebbles such that there is solution by moving each blue pebble at most one step if and only if $G$ has a dominating set of size $k$. Let $v_1, \ldots, v_n$ be the vertices of $G$. We construct a graph $F$ as follows. We start with vertices $s$, $a_1, \ldots, a_n$, $b_1, \ldots, b_n$ where $s$ is connected to every $a_i$, and the $a_i$'s form a clique of size $n$. Furthermore, if $v_i$ and $v_j$ are neighbors in $G$, then $a_i$ and $b_j$ are adjacent in $F$. We place $k$ blue pebbles on $s$ and one red pebble on each $b_i$.

If $G$ has a dominating set $v_{i_1}, \ldots, v_{i_k}$, then we move the $k$ blue pebbles to $a_{i_1}, \ldots, a_{i_k}$. Note that the $k$ blue pebbles induce a clique in $F$. Furthermore, if vertex $v_j$ of $G$ is dominated by its neighbor $v_{i_\ell}$, then the red pebble on vertex $b_j$ is adjacent to the blue pebble on vertex $a_{i_\ell}$. Thus the red pebbles are adjacent to the clique induced by the blue pebbles, hence the graph induced by all the pebbles is connected. The other direction is also easy to see: if the facility pebbles move to vertices $a_{i_1}, \ldots, a_{i_k}$ such that the red and blue pebbles together induce a connected graph, then vertices $v_{i_1}, \ldots, v_{i_k}$ form a dominating set in $G$. □

In planar graphs, however, STEINER CONNECTIVITY is FPT parameterized by the number of connector (blue) pebbles. We only sketch the proof, which is a quite simple bounded search tree algorithm, using a combinatorial observation on the structure of planar graphs.

**Theorem 8.6.** STEINER CONNECTIVITY, *parameterized by the number of connector pebbles, is* FPT *on planar graphs.*

*Proof.* First, contracting each connected component of red pebbles to a single vertex does not change the problem, hence we can assume that the red pebbles are independent. It follows that the blue pebbles dominate the red pebbles in the solution, i.e., each red pebble has a blue neighbor. Let $k$ be the number of blue pebbles. If there is no vertex having more than $2k^2$ red neighbors, then the $k$ blue pebbles can dominate at most $2k^3$ red pebbles, i.e., it can be assumed that there is only a bounded number of red pebbles and the algorithm in Section 3.1 can be applied. Suppose that some vertex $v$ has more than $2k^2$ red neighbors.

Then either a blue pebble is moved to $v$ in a solution, or the red vertices are dominated some other way, implying that there is a vertex $u$ having more than $2k^2/k = 2k$ red neighbors common with $v$. By planarity, every vertex other than $u$ and $v$ can be adjacent to at most two of these red pebbles, hence they cannot be dominated if neither $u$ and $v$ is used. Thus we can branch into $2k$ directions: one of the $k$ blue pebbles has to be moved to one of $u$ and $v$. Formally, moving a blue pebble $p$ to, say, $u$ means changing the initial location of $p$ to $u$ and making it stationary by making the cost of moving $p$ to any other vertex infinite. Since pebble $p$ cannot move and $u$ has a red neighbor, changing the color of $p$ to red does not modify the problem. Thus in each branching step, we decrease the number blue pebbles, which means that the search tree has height at most $k$ and has at most $2^k$ leaves. □

## 8.4 Geometric Graphs

*Proof (of Theorem 4.2):* There are at most $2^{k^2}$ possible graphs on $k$ vertices. Suppose we are given a list of all possible allowed subgraphs. We guess which graph in this subset will be obtained by the pebbles (there are at most $2^{k^2}$ choices). We guess the mapping of which pebbles will be mapped to which vertices of the intended graph (there are at most $k^k$ choices). It remains to optimize the maximum or total movement to realize this solution.

We use Canny's Roadmap Algorithm for general Euclidean motion planning [5]. Given $m$ semi-algebraic constraints, each of degree $\delta$, over $r$ real variables, this (randomized) algorithm computes a connectivity-preserving representation of the space of all solutions in $m^{O(r)}\delta^{O(r^2)}$ expected time. (Alternatively, the deterministic version of the algorithm runs in $m^{O(r)}\delta^{O(r^4)}$; and a new deterministic algorithm runs in $m^{O(r)}\delta^{O(r^{1.5})}$ [2].) In particular, the algorithm decides whether the space is nonempty.

We use the Roadmap Algorithm as follows. For each vertex $i$, we make $d$ real variables representing the point $p_i$ to which that agent moves. For each edge $(i,j)$ present in the graph, we add the semi-algebraic constraint $\|p_i - p_j\|^2 \leq 1$ (or $< 1$). For each edge $(i,j)$ absent from the graph, we add the semi-algebraic constraint $\|p_i - p_j\|^2 > 1$ (or $\geq 1$). We also add the algebraic constraints $p_i = o_i + v_i$, where $o_i$ is the known (constant) original position and $v_i$ is a vector of $d$ additional variables representing the motion of agent $i$, and we add the algebraic constraint $m_i^2 = \|v_i\|^2$, where $m_i$ is another real variable to measure the motion. To decide whether the total motion can be at most $x$, we can add a constraint $m_1 + m_2 + \cdots + m_n \leq x$ and test with the Roadmap Algorithm whether the space is nonempty. To decide whether the maximum motion can be at most $x$, we can add $n$ constraints $m_i \leq x$ and test with the Roadmap Algorithm whether the space is nonempty. In total, we have $O(kd)$ real variables and $O(k^2 + kd)$ semi-algebraic constraints, so the Roadmap Algorithm takes $(k^2 + kd)^{kd} 2^{O((kd)^2)} = 2^{O((kd)^2)}$ expected time.

Finally we binary search on the objective. Clearly the optimal solution we seek has cost at least 0. We also claim that it has cost at most $2k^2 + kD$: move all agents to a common point, which costs at most $kD$; and then move the agents to a modified instance of the target graph where every connected component (which has diameter at most $k$) has distance at most 2 to another connected component, and thus the overall diameter is at most $2k$, for a cost of at most $2k^2$. Thus a binary search, terminating when we have an interval of length at most $\varepsilon$ (and thus have an additive $\varepsilon$ approximation), requires $O(\lg[(2k^2 + kD)/\varepsilon]) = O(\lg k + \lg D + \lg(1/\varepsilon))$ calls to the Roadmap Algorithm.

The overall expected running time is $2^{k^2} k^k \cdot 2^{O((kd)^2)} \cdot (\lg k + \lg D + \lg(1/\varepsilon)) = 2^{O((kd)^2)}(\lg D + \lg(1/\varepsilon))$. □

## 8.5 Improving CONNECTIVITY with Fast Subset Convolution

Here we consider the version of CONNECTIVITY where the pebbles have to be moved such that they induce a connected graph, and it is allowed that more than one pebbles are on the same vertex. We have seen in Section 3.1 that the general results of Theorem 3.1 imply that this problem is FPT. Here we give a more efficient algorithm for this specific problem: the exponential factor of the running time is only $O(2^k)$. The main tool we use is the recent algorithm of Björklund et al. [3] for fast subset convolution in the min-sum semiring:

**Theorem 8.7** (Björklund et al. [3])**.** *Let $P$ be a set of size $k$ and let $f, g : 2^N \to \{-M, \ldots, M\}$ be two functions. The $2^k$ values of the function*

$$h(S) := \min_{T \subseteq S}(f(T) + g(S \setminus T))$$

can be evaluated in time $\tilde{O}(2^k M)$.

(The notation $\tilde{O}$ suppresses polylogarithmic factors.) For each vertex $v$ and subset $S \subseteq P$ of pebbles, let $f_v(S)$ (resp., $f'_v(S)$) be the minimum cost of moving the pebbles in $S$ to induce a connected subgraph including $v$ (resp., including a *neighbor* of $v$). Let $f_{v,i}(S)$ be $f_v(S)$ if $|S| \le i$ and infinity otherwise and define $f'_{v,i}(S)$ similarly. Observe that $f_{v,1}(S)$ is trivial to determine and if $f_{v,i}(S)$ is known for every $v$, then $f'_{v,i}(S)$ is easy to determine as well (as $f'_{v,i}(S)$ is the minimum of $f_{u,i}(S)$, taken over all neighbors $u$ of $v$).

Our strategy is to compute, iteratively for $i = 1, 2, ..., k$, the value of $f_{v,i}(S)$ (and hence the value of $f'_{v,i}(S)$) for every $v$ and $S$. To determine $f_{v,i}(S)$, first we have to consider the trivial case when all the pebbles are moved to $v$; the cost of this is easy calculate. Otherwise, if $F$ is the connected subgraph induced by the pebbles in $S$ (after the movement), then $F \setminus v$ has at least one connected component, say $C$, and this component $C$ contains a nonempty proper subset $T \subset S$ of the pebbles. In this case, the cost of the movement can be expressed as $f'_{v,i}(T) + f_{v,i}(T \setminus S) = f'_{v,i-1}(T) + f_{v,i-1}(S \setminus T)$ (the equality follows from $|T| < |S| \le i$, $|S \setminus T| < |S| \le i$). Thus we have to determine $\min_{T \subseteq S}(f'_{v,i-1}(T) + f_{v,i-1}(S \setminus T))$, and the algorithm of Theorem 8.7 can be used for this purpose (as the functions $f'_{v,i-1}$ and $f_{v,i-1}$ were already determined). If the values of the movement cost function are polynomially bounded, then the resulting algorithm has running time $O(2^k n^{O(1)})$.

**Theorem 8.8.** *The movement problem* CONNECTIVITY *with $k$ pebbles can be solved in time $2^k \cdot n^{O(1)}$ if the values of the movement cost function are polynomially bounded.*

## 8.6 Hereditary Properties

To prove Theorem 3.3 we use the following hardness result on the parameterized complexity of finding induced subgraphs with hereditary properties. Given a graph property $\mathcal{H}$, in the problem INDUCED-$\mathcal{H}$, we are given a graph $G$ and an integer $k$, and the task is to find $k$ vertices of $G$ that induce a subgraph in $\mathcal{H}$.

**Theorem 8.9** (Khot and Raman [24])**.** *Let $\mathcal{H}$ be a decidable hereditary graph property. If $\mathcal{H}$ includes all empty and all complete graphs, then* INDUCED-$\mathcal{H}$ *is* FPT*, and W[1]-hard otherwise.*

*Proof (of Theorem 3.3).* Let $\mathcal{G}$ be a hereditary property where each vertex has exactly one client agent and there are no other type of pebbles. Let $\mathcal{G}_0$ contain the underlying graphs of the multi-colored graphs in $\mathcal{G}$; clearly $\mathcal{G}_0$ is hereditary. Let $\mathcal{G}'_0$ contain all the graphs in $\mathcal{G}_0$ and those graphs $G$ that can be made a member of $\mathcal{G}_0$ by connecting an arbitrary vertex of $G$ with every other vertex. For example, if $\mathcal{G}_0$ is the set of all cliques, then $\mathcal{G}'_0$ contains those graphs where there is only at most one vertex which is not connected to every other vertex. On the other hand, if $\mathcal{G}_0$ is the set of all independent sets, then $\mathcal{G}'_0 = \mathcal{G}_0$. It is easy to see that $\mathcal{G}'_0$ is hereditary, the largest clique appearing in $\mathcal{G}'_0$ is the same as the largest clique appearing in $\mathcal{G}_0$, and the size of the largest independent set in $\mathcal{G}'_0$ is at most one larger than the size of the largest independent set in $\mathcal{G}_0$. Therefore, INDUCED-$\mathcal{G}'_0$ is W[1]-hard. We reduce this problem to the movement problem.

To prove the hardness of the movement problem, we reduce an instance $(G, k)$ of INDUCED-$\mathcal{G}'_0$ to the movement problem. Given a graph $G$, we construct a graph $G'$ by adding a new vertex $v$ that is adjacent to every other vertex. For every possible combinations of $k$ colors, we put $k$ pebbles with these colors on $v$ and check whether the movement problem has a solution where each pebble moves at most one step. We claim that the movement problem has such a solution for at least one combination of colors if and only if $G$ has an induced subgraph $H \in \mathcal{G}'_0$ on $k$ vertices.

Assume first that the movement problem has a solution. We consider two cases. If all the pebbles move, then they induce a multicolored graph that belongs to $\mathcal{G}$. The underlying graph of this multicolored graph in $\mathcal{G}$ is in $\mathcal{G}_0 \subseteq \mathcal{G}'_0$ by definition, that is, $G$ has an induced subgraph that belongs to $\mathcal{G}'_0$. If not all pebbles move, then exactly one pebble $p$ stays at $v$ (as $\mathcal{G}$ has no member with more than one pebble at a vertex). The $k$ pebbles induce a graph $H \in \mathcal{G}_0$ and by construction $p$ is adjacent to every other pebble, i.e., $H$ has a universal vertex. Thus if we move pebble $p$ to an arbitrary vertex $u$ not occupied by the other pebbles, then the $k$ pebbles induce a graph in $\mathcal{G}'_0$ (as this graph can be made a member of $\mathcal{G}_0$ by making $u$ an universal vertex).

The other direction is also easy to see. If $G$ has a subgraph $H \in \mathcal{G}_0 \subseteq \mathcal{G}'_0$, then by moving appropriately colored pebbles to these vertices we can obtain a solution where the $k$ pebbles induce a multicolored graph whose underlying graph is $H$. Assume now that $G$ has a subgraph $H \in \mathcal{G}'_0 \setminus \mathcal{G}_0$, i.e., connecting vertex $u \in H$ with every other vertex gives a graph $H' \in \mathcal{G}_0$. In this case we can obtain a solution by moving $k-1$ appropriately colored pebbles to the $k-1$ vertices corresponding to $H \setminus u$ and leaving one pebble at $v$. □

## 9 Conclusions

We have introduced a very general formulation of movement minimization problems and investigated their fixed-parameter tractability, parameterized by the number of pebbles (of certain type). We obtained general meta-theorems, both on the algorithmic and complexity side, which give us convenient tools to classify concrete problems. The interesting feature of this results is that they reduce an algorithmic/complexity question (is the problem FPT?) to a purely combinatorial question (is treewidth bounded?). By looking more closely at certain variants, we observed that a wide range of algorithmic ideas are relevant for this class of problems. Nevertheless, there are many natural problem variants for which our results do not give satisfying answers. We list a few concrete questions to stimulate further research:

- Given a set of pebbles in a graph $G$, move the pebbles such that they induce a connected graph and the total movement is at most $k$ steps. Is this problem FPT in general graphs/planar graphs, parameterized by $k$?

- For which problems can we use bidimensionality theory to obtain subexponential FPT algorithms (i.e., running time of the form $2^{o(k)} n^{O(1)}$)?

- Prove that natural movement problems are NP-hard in the geometric setting. For example, in the DISPERSION problem (move the agents such that the distance between any two of them is at least 1), is it NP-hard to minimize the sum or the maximum of the movements?

Finally, let us mention that the general hardness proof of Theorem 3.2 is limited in the sense that it proves hardness only for a very specific question: finding a solution where every pebble moves at most one step. While intuitively every reasonable question should be at least as hard as this one-step question, this hardness result does not formally rule out the possibility that some other natural question is FPT. For example, it can happen that the one-step question is W[1]-hard, but minimizing the total movement is FPT. It would be interesting to adapt the hardness result to other reasonable questions.

## References

[1] N. Alon, R. Yuster, and U. Zwick. Color-coding. *J. ACM*, 42(4):844–856, 1995.

[2] S. Basu, M.-F. Roy, M. S. E. Din, and É. Schost. A baby step-giant step roadmap algorithm for general algebraic sets. 2012. http://arxiv.org/abs/1201.6439.

[3] A. Björklund, T. Husfeldt, P. Kaski, and M. Koivisto. Fourier meets Möbius: fast subset convolution. In *STOC*, pages 67–74, New York, NY, USA, 2007. ACM.

[4] J. L. Bredin, E. D. Demaine, M. Hajiaghayi, and D. Rus. Deploying sensor networks with guaranteed capacity and fault tolerance. In *MOBIHOC*, pages 309–319, Urbana-Champaign, Illinois, May 25–28 2005.

[5] J. F. Canny. *The Complexity of Robot Motion Planning*. MIT Press, 1987.

[6] P. Corke, S. Hrabar, R. Peterson, D. Rus, S. Saripalli, and G. Sukhatme. Autonomous deployment of a sensor network using an unmanned aerial vehicle. In *ICRA*, New Orleans, USA, 2004.

[7] P. Corke, S. Hrabar, R. Peterson, D. Rus, S. Saripalli, and G. Sukhatme. Deployment and connectivity repair of a sensor net with a flying robot. In *ISER*, Singapore, 2004.

[8] B. Courcelle. Graph rewriting: an algebraic and logic approach. In *Handbook of Theoretical Computer Science, Vol. B*, pages 193–242. Elsevier, Amsterdam, 1990.

[9] E. D. Demaine and M. Hajiaghayi. The bidimensionality theory and its algorithmic applications. *The Computer Journal*, 51(3):292–302, 2008.

[10] E. D. Demaine, M. Hajiaghayi, and K. Kawarabayashi. Algorithmic graph minor theory: Decomposition, approximation, and coloring. In *FOCS 2005*, pages 637–646, 2005.

[11] E. D. Demaine, M. T. Hajiaghayi, H. Mahini, A. S. Sayedi-Roshkhar, S. O. Gharan, and M. Zadimoghaddam. Minimizing movement. *ACM Transactions on Algorithms*, 5(3), 2009.

[12] S. Doddi, M. V. Marathe, A. Mirzaian, B. M. E. Moret, and B. Zhu. Map labeling and its generalizations. In *SODA*, pages 148–157, New Orleans, LA, 1997.

[13] R. G. Downey and M. R. Fellows. *Parameterized Complexity*. Monographs in Computer Science. Springer, New York, 1999.

[14] D. Eppstein. Subgraph isomorphism in planar graphs and related problems. *J. Graph Algorithms Appl.*, 3(3), 1999.

[15] M. R. Fellows and H. Fernau. Facility location problems: A parameterized view. *Discrete Applied Mathematics*, 159(11):1118 – 1130, 2011.

[16] J. Flum and M. Grohe. *Parameterized Complexity Theory*. Springer-Verlag, Berlin, 2006.

[17] E. C. Freuder. Complexity of $k$-tree structured constraint satisfaction problems. In *AAAI*, pages 4–9, 1990.

[18] Z. Friggstad and M. R. Salavatipour. Minimizing movement in mobile facility location problems. In *49th Annual IEEE Symposium on Foundations of Computer Science (FOCS'08)*, pages 357–366, 2008.

[19] M. Grohe. The complexity of homomorphism and constraint satisfaction problems seen from the other side. *J. ACM*, 54(1):1, 2007.

[20] T.-R. Hsiang, E. M. Arkin, M. A. Bender, S. P. Fekete, and J. S. B. Mitchell. Algorithms for rapidly dispersing robot swarms in unknown environments. In *Algorithmic Foundations of Robotics V*, volume 7 of *Springer Tracts in Advanced Robotics*, pages 77–94. Springer-Verlag, 2003.

[21] F. Hüffner, R. Niedermeier, and S. Wernicke. Techniques for practical fixed-parameter algorithms. *The Computer Journal*, 51(1):7–25, 2008.

[22] M. Jiang, S. Bereg, Z. Qin, and B. Zhu. New bounds on map labeling with circular labels. In *ISAAC*, volume 3341 of *Lecture Notes in Computer Science*, pages 606–617, Hong Kong, China, December 2004.

[23] M. Jiang, J. Qian, Z. Qin, B. Zhu, and R. Cimikowski. A simple factor-3 approximation for labeling points with circles. *Information Processing Letters*, 87(2):101–105, 2003.

[24] S. Khot and V. Raman. Parameterized complexity of finding subgraphs with hereditary properties. *Theoret. Comput. Sci.*, 289(2):997–1008, 2002.

[25] T. Kloks. *Treewidth*, volume 842 of *Lecture Notes in Computer Science*. Springer-Verlag, Berlin, 1994.

[26] S. M. LaValle. *Planning Algorithms*. Cambridge University Press, 2006. http://msl.cs.uiuc.edu/planning/.

[27] R. Niedermeier. *Invitation to fixed-parameter algorithms*, volume 31 of *Oxford Lecture Series in Mathematics and its Applications*. Oxford University Press, Oxford, 2006.

[28] J. Plehn and B. Voigt. Finding minimally weighted subgraphs. In *Graph-theoretic concepts in computer science (Berlin, 1990)*, volume 484 of *Lecture Notes in Comput. Sci.*, pages 18–29. Springer, Berlin, 1991.

[29] J. H. Reif and H. Wang. Social potential fields: a distributed behavioral control for autonomous robots. In *Proceedings of the Workshop on Algorithmic Foundations of Robotics*, pages 331–345, 1995.

[30] N. Robertson, P. D. Seymour, and R. Thomas. Quickly excluding a planar graph. *J. Combin. Theory Ser. B*, 62(2):323–348, 1994.

[31] A. C. Schultz, L. E. Parker, and F. E. Schneider, editors. *Multi-Robot Systems: From Swarms to Intelligent Automata.* Springer, 2003. Proceedings from the 2003 International Workshop on Multi-Robot Systems.

[32] T. Strijk and A. Wolff. Labeling points with circles. *International Journal of Computational Geometry & Applications*, 11(2):181–195, 2001.

# A  Proof of Theorem 6.2

Let $v_1, \ldots, v_k$ be the vertices of $F$, and assign labels $\{1, \ldots, k\}$ to the vertices of $G$ uniformly and independently at random. We say that a mapping $\phi$ from a subset $S \subseteq V(F)$ to $V(G)$ is *colorful* if the vertices in $\phi(S)$ have distinct labels. Below we present an algorithm for finding the minimum cost colorful subgraph embedding with respect to a given labeling. This cost might be larger than the cost of the minimum cost embedding. However, an arbitrary embedding is colorful in a random labeling with probability at least $k!/k^k = 2^{-O(k)}$, hence if we try $2^{O(k)}$ random labelings, then with constant probability at least one optimum embedding becomes colorful in one of the labelings. This means that the probability of not finding the optimum can be reduced to an arbitrarily small constant by trying $2^{O(k)}$ random labelings. The algorithm can be derandomized by using a $k$-perfect family of hash functions instead of the random labelings [1], [16, Section 13.3].

Let the rooted tree $T$ with the bags $(B_u \subseteq V(G) : u \in V(T))$ be a width $t$ tree decomposition of $F$. We assume this decomposition is *nice* [25]: every node $u \in V(T)$ is either a *leaf node* ($u$ has no children), *join node* ($u$ has two children $u', u''$ and $B_{u'} = B_{u''}$), *forget node* ($u$ has a single child $u'$ and $B_u \subseteq B_{u'}$, $|B_u| = |B_{u'}| - 1$), or *introduce node* ($u$ has a single child $u'$ and $B_u \supseteq B_{u'}$, $|B_u| = |B_{u'}| + 1$). For every $u \in V(T)$, we denote by $V_u$ the union of the bags $B_{u'}$ for every descendant $u'$ of $u$ (including $u$ itself). Let us fix a labeling $\ell : \{1, \ldots, k\} \to V(G)$ of the vertices. For every $u \in V(T)$, $L \subseteq \{1, \ldots, k\}$, and mapping $\psi : B_u \to V(G)$, we define $c(u, L, \psi)$ to be the cost of the minimum cost subgraph embedding $\phi : V_u \to V(G)$ of $G[V_u]$ in $G$ satisfying

1. $\phi$ is colorful,
2. $\phi(v) = \psi(v)$ for every $v \in B_u$, and
3. $\phi(V_u)$ uses only the labels in $L$.

If there is no mapping $\phi$ satisfying the requirements, then $c(u, L, \psi)$ is defined to be $\infty$. It follows from these definitions that if $r$ is the root of $T$, then the minimum cost of a colorful embedding of $F$ into $G$ is given by $\min_\psi c(r, \{1, \ldots, k\}, \psi)$, where the minimum is taken over all mappings $\psi : B_r \to V(G)$.

We determine the values $c(u, L, \psi)$ in a bottom up traversal of the tree $T$, i.e., we assume that these values are already determined for every child of $u$. Depending the type of node $u \in V(F)$ we do the following.

- *Node $u$ is a leaf node.* In this case $V_u = B_u$. Thus we can simply try all subgraph embeddings $\phi : B_u \to V(G)$ of $G[B_u]$ in $G$ and define $c(u, L, \psi)$ to be the minimum cost of a mapping $\phi$ that satisfies all three constraints above.

- *Node $u$ is an introduce node with child $u'$.* Let $B_u \setminus B_{u'} = \{v\}$. To compute the value $c(u, L, \psi)$ for a given $L$ and $\psi$, we first check if $\ell(\psi(v)) \in L$; if not, then clearly $c(u, L, \psi) = \infty$. Next we check if for every neighbor $w \in B_u$ of $v$, $\psi(w)$ is a neighbor of $u$. If this does not hold for some $w$, then again we have $c(u, L, \psi) = \infty$. Otherwise, let $\psi'$ be $\psi$ restricted to $B_{u'}$. Now it is easy to see that $c(u, L, \psi) = c(u', L \setminus \{\ell(\psi(v))\}, \psi') + c(v, \psi(v))$: any mapping $\phi'$ realizing the minimum $c(u', L \setminus \{\ell(\psi(v))\}, \psi')$ can be extended to a suitable mapping $\phi$ by defining $\phi(v) = \psi(v)$.

- *Node $u$ is a forget node with child $u'$.* Let $B_{u'} \setminus B_u = \{v\}$. It is easy to see that $c(u, L, \psi)$ is the minimum of $c(u', L, \psi')$, taken over every $\psi' : B_{u'} \to V(G)$ satisfying $\psi'(w) = \psi(w)$ for every $w \in B_u$. Thus $c(u, L, \psi)$ can be obtained as the minimum of $n$ already determined values.

- *Node $u$ is a join node with children $u'$ and $u''$.* Let $L_0$ be the labels of the vertices $\psi(B_u)$ and let $C_u := \sum_{v \in B_u} c(v, \psi(v))$. We show that

$$c(u, L, \psi) = \min_{\substack{L', L'' \subseteq L \\ L' \cap L'' = L_0}} \big(c(u', L', \psi) + c(u'', L'', \psi) - C_u\big), \quad (1)$$

  which means that $c(u, L, \psi)$ can be determined in time $2^{O(k)}$ using values already determined. Suppose that for some $L', L'' \subseteq L$ with $L' \cap L'' = L_0$, mappings $\phi' : V_{u'} \to V(G)$ and $\phi'' : V_{u''} \to V(G)$ are minimum cost subgraph embeddings that define the values $c(u, L', \psi')$ and $c(u, L'', \psi'')$, respectively. Since $\psi'$ and $\psi''$ are the same as $\psi$ on $B_u$, they can be joined to obtain a subgraph embedding $\phi : V_u \to V(G)$. Note that $\psi$ is injective: if $\phi'(w') = \phi''(w'')$, then $\ell(\phi'(w)) \in L' \cap L'' = L_0$, which implies that $w', w'' \in B_u$ and $w' = w''$ follows. The cost of $\phi$ is $\sum_{v \in V_{u'}} c(v, \phi'(v)) + \sum_{v \in V_{u''}} c(v, \phi''(v)) - C_u$, where the last term accounts for the double counting of the vertices in $B_u$. Thus $c(u, L, \psi)$ cannot be larger than the right hand side in (1).

  Conversely, if $\phi$ is an embedding of $V_u$ that defines $c(u, L, \psi)$, then let $L' := \ell(\psi(V_{u'}))$ and $L'' := \ell(\psi(V_{u''}))$. Since $V_{u'} \cap V_{u''} = B_u$, $L' \cap L'' = \phi(B_u) = \psi(B_u) = L_0$. Restricting $\phi$ to $V_{u'}$ and $V_{u''}$ gives two embeddings $\phi'$ and $\phi''$, respectively. By considering these two particular embeddings $\phi', \phi''$, it is easy to see that the right hand side of (1) is at most $c(u, L, \psi)$, proving the equality.

As discussed above, a value $c(u, L, \psi)$ can be determined in time $2^{O(k)} \cdot n^{O(t)}$ if the values corresponding to the children of $u$ are already determined. As there are $2^{O(k)} \cdot n^{O(t)}$ values $c(u, L, \psi)$ and the running time of every other part of the algorithm (such as finding the tree decomposition) is dominated by $2^{O(k)} \cdot n^{O(t)}$, the total running time is $2^{O(k)} \cdot n^{O(t)}$.